%% file: main.tex
\documentclass[journal]{vgtc}                % final (journal style)
\ifpdf%                                % if we use pdflatex
  \pdfoutput=1\relax                   % create PDFs from pdfLaTeX
  \usepackage{graphicx}                % allow us to embed graphics files
  \DeclareGraphicsExtensions{.pdf,.png,.jpg,.jpeg} % for pdflatex we expect .pdf, .png, or .jpg files
\else%                                 % else we use pure latex
  \usepackage{graphicx}                % allow us to embed graphics files
  \DeclareGraphicsExtensions{.eps}     % for pure latex we expect eps files
\fi%

\graphicspath{{figures/}{pictures/}{images/}{./}} % where to search for the images

\usepackage{microtype}                 % use micro-typography (slightly more compact, better to read)
\PassOptionsToPackage{warn}{textcomp}  % to address font issues with \textrightarrow
\usepackage{textcomp}                  % use better special symbols
\usepackage{mathptmx}                  % use matching math font
\usepackage{times}                     % we use Times as the main font
\usepackage{cite}                      % needed to automatically sort the references
\usepackage{tabu}                      % only used for the table example
\usepackage{booktabs}                  % only used for the table example
\usepackage{paralist}
\usepackage{amsmath}
\usepackage{balance}
\usepackage[T1]{fontenc}

\usepackage[normalem]{ulem}

\onlineid{1234}

\vgtccategory{Research}
\vgtcpapertype{application/design study}
\title{QLens: Visual Analytics of Multi-step Problem-solving \\ Behaviors for Improving Question Design}

\shortauthortitle{Biv \MakeLowercase{\textit{et al.}}: Global Illumination for Fun and Profit}

\author{Meng Xia, Reshika Palaniyappan Velumani, Yong Wang, Huamin Qu, and Xiaojuan Ma}
\authorfooter{
\item
 M. Xia, R.P. Velumani, Y. Wang, H. Qu, and X. Ma are with Hong Kong University of Science and Technology. Email: $iris.xia$@connect.ust.hk, $\{$reshikapv, jiayouwyhit$\}$@gmail.com, $\{$huamin, mxj$\}$@cse.ust.hk
}

\abstract{
With the rapid development of online education in recent years, there has been an increasing number of learning platforms that provide students with multi-step questions to cultivate their problem-solving skills.
To guarantee the high quality of such learning materials, question designers need to inspect how students' problem-solving processes unfold step by step to infer whether students' problem-solving logic matches their design intent. They also need to compare the behaviors of different groups (e.g., students from different grades) to distribute questions to students with the right level of knowledge.
The availability of fine-grained interaction data, such as mouse movement trajectories from the online platforms, provides the opportunity to analyze problem-solving behaviors. However, it is still challenging to interpret, summarize, and compare the high dimensional problem-solving sequence data.
In this paper, we present a visual analytics system, \textit{QLens}, to help question designers inspect detailed problem-solving trajectories, compare different student groups, distill insights for design improvements. 
In particular, \textit{QLens} models problem-solving behavior as a hybrid state transition graph and visualizes it through a novel glyph-embedded Sankey diagram, which reflects students' problem-solving logic, engagement, and encountered difficulties. 
We conduct three case studies and three expert interviews to demonstrate the usefulness of \textit{QLens} on real-world datasets that consist of thousands of problem-solving traces.
} % end of abstract


\CCScatlist{ % not used in journal version
 \CCScat{K.6.1}{Management of Computing and Information Systems}%
{Project and People Management}{Life Cycle};
 \CCScat{K.7.m}{The Computing Profession}{Miscellaneous}{Ethics}
}

\teaser{
  \centering
  \includegraphics[width=\linewidth]{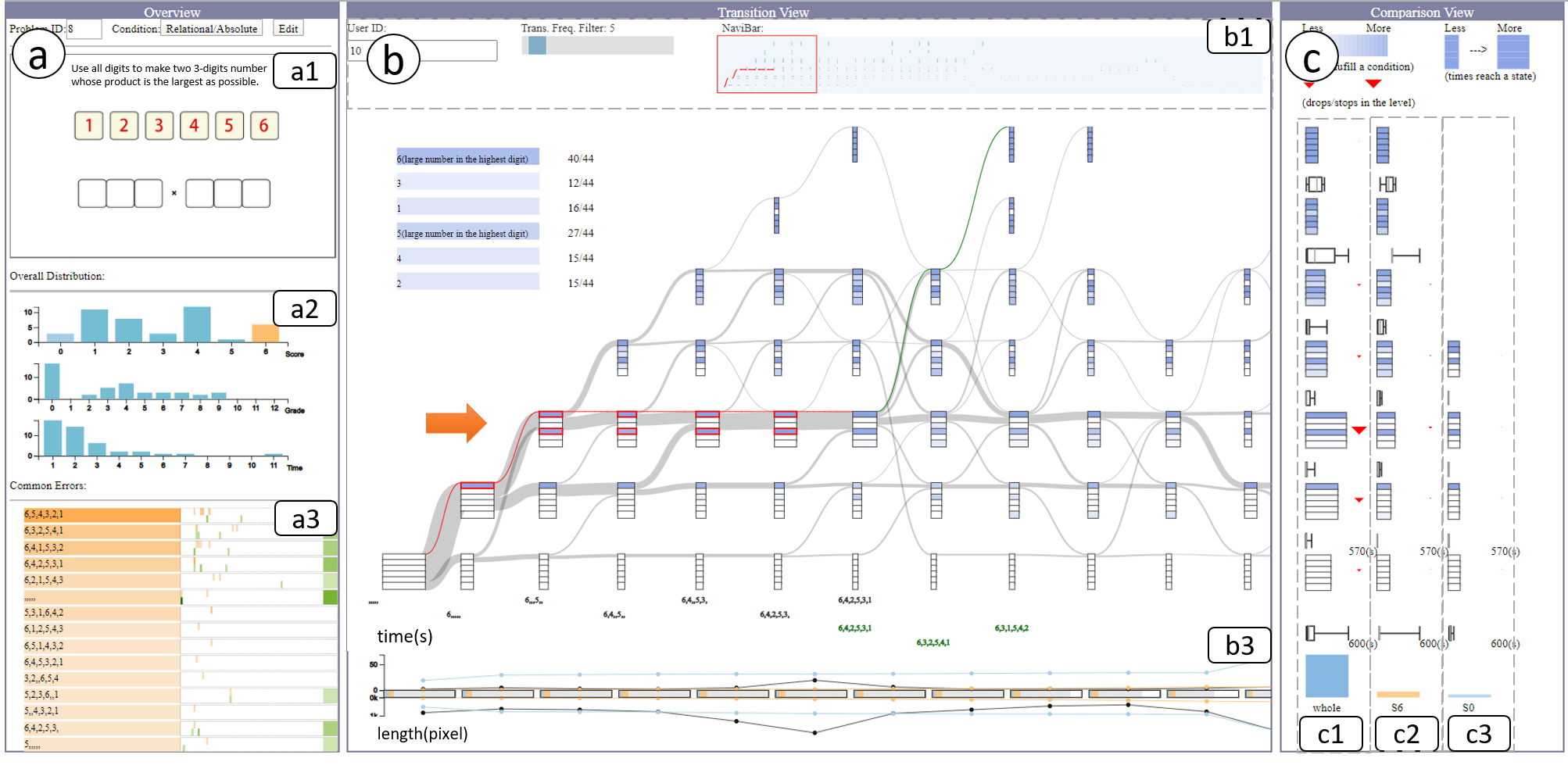}
  \caption{\textit{QLens} enables question designers to analyze students' multi-step problem-solving behaviors for design improvements from levels of detail. (1) Macro-level: Overview (a) shows the question's preview (a1), students' overall performance (a2) and the ranking of common wrong answers based on the frequency of occurence(a3). (2) Meso-level: Transition View (b) visualizes the problem-solving processes to reflect how a group of students proceed step by step (\textit{problem-solving logic}) using a novel glyph-embedded Sankey diagram (b2) and the amount of efforts (\textit{engagement}) using a contextual axis (b3). In addition, Comparison  View (c) enables users to compare the problem-solving logic, engagement, and encountered difficulties of different groups (c1, c2, c3). (3) Micro-level (the highlighted path in b): typical incorrect paths and the corresponding data-driven recommended paths are demonstrated for question designers to evaluate the feasibility of the data-driven feedback. Rich interactions are also provided for exploration (e.g., filters in b1).
}
	\label{fig:teaser}
}

\vgtcinsertpkg

\begin{document}
\maketitle
\input{sections/01-introduction.tex}

\input{sections/02-relatedwork.tex}
\input{sections/03-system.tex}

\input{sections/04-model.tex}
\input{sections/05-visualization.tex}
\input{sections/06-evaluation.tex}

\input{sections/07-discussion.tex}

\input{sections/08-conclusion.tex}

%% if specified like this the section will be committed in review mode
\acknowledgments{
This research is partially supported by Theme-based Research Scheme of the Hong Kong RGC under grant T44-707/16-N. We would like to thank our industry sponsor, Trumptech (Hong Kong) Ltd, for providing with the data and the research platform.}

\bibliographystyle{abbrv-doi}

\bibliography{main}
\end{document}

%% file: sections/01-introduction.tex
\section{Introduction}

\textcolor{black}{With online education becoming increasingly popular in the past decades, various types of online learning materials including multi-step questions are provided for students to cultivate their problem-solving skills. 
% A step is the smallest user interface interaction for which it makes sense to call it correct or incorrect~\cite{vanlehn2016regulative}. 
% Multi-step problems are those which require students to conduct interactions for more than one step. There are different types of multi-step problems, e.g., problems with multiple blanks to fill, problems with multiple elements to drag, multiple-choice questions.
% \yong{Meng, the original logic here is quite abrupt and these sentences come in a sudden.}
For example,
% various MOOC platforms (e.g., Khan Academy),
online question pools (e.g., LeetCode~\cite{LeetCode}, LearnLex~\cite{LearnLex}), and intelligent tutoring systems (e.g., SimStudent~\cite{SimStudent}) offers interactive maths questions and/or programming exercises ~\cite{xia2019peerlens}. 
Different from traditional multiple-choice questions, these newly-designed questions require students to construct a solution that fulfills a series of conditions by conducting multi-step interactions~\cite{glassman2013toward}, which are called \textit{multi-step questions}.
A step is \textit{``the smallest user interface actions for which it makes sense to call it correct
or incorrect''}~\cite{vanlehn2016regulative}. 
For example, interactions (e.g., drag-and-drops) that change the answer of the question is a step, but mouse movements without changing the answer is not regarded as a step.
% For multi-step questions, one step refers to the smallest user interaction that can influence the judgement of whether a question is correctly answered or not~\cite{vanlehn2016regulative}.
% \yong{Meng, pls double check if this definition is correct or not.}
% Multi-step problems are those which require students to conduct interactions for more than one step. 
There are different types of multi-step questions, for example, questions with multiple blanks to fill, questions with multiple elements to drag.
% multiple-choice questions.
Such questions are becoming popular learning materials practised by tens of thousands of students~\cite{xia2019peerlens}.
}

% inspect students' detailed problem-solving behaviors and check whether the design intention is followed to improve the question design (e.g, adjusting the setting of testing cases). In particular, they require to understand how students' detailed problem-solving processes unfold step by step and to compare behaviors of different groups of students (e.g., grades).
% Few studies have explored the problem-solving behaviors at a fine-grained level to adequately support analytical tasks such as interpretation, summarization, and comparison. 

% The question description, testing cases, solution provided, and the feedback design (e.g., hints) are all needed to be considered by question designers to guarantee high-quality learning material~\cite{??}. 

% 2 Research problem/user task this work
% By linking 
% what students are doing on a fine-grained level with what educators are expecting students to do may 
% help to inform how to improve the learning material design~\cite{nguyen2018students, ??}. 
% To guarantee a higher quality learning material, inspect students' detailed problem-solving behaviors and check whether the design intention is followed by students to improve the question design. For example, question designers need to consider the clarity of the question description, the setting of testing cases, the comprehensiveness of the solution provided, the target group of the question, and the feasibility of the data-driven feedback (e.g., hints).

Prior studies~\cite{nguyen2018student, glassman2013toward} have shown that linking students' fine-grained problem-solving behaviors with the educators' expectation of students' performance can improve the design of learning materials (e.g., online questions). 
Problem-solving behaviors are a sequence of steps that a student takes to produce the final answer~\cite{piech2015autonomously}.
Exploring the patterns emerged from thousands of such processes can provide deep insights in 
understanding students' problem-solving process, and a set of crucial questions can also be answered. For example,
What approaches do the students take? How much effort do students spend on solving a problem? What kind of difficulties do students commonly encounter in their practices?
By knowing why and how students may fail on a question,
educators and question designers are able to adjust
the learning designs (e.g., the setting of testing cases), distribute questions to students with the right level of knowledge, and provide students with suitable on-the-fly guidance.

% the clarity of the question description, the setting of testing cases, the comprehensiveness of the solution provided, the target group of the question, and the feasibility of the data-driven feedback (e.g., hints).

% 3 Very related work + your difference from the previous work
Existing research on analyzing problem-solving behaviors mainly focus on the outer loop (i.e., how students master the knowledge by solving a series of problems)~\cite{liu2017going, vanlehn2006behavior}.
But these methods only took the final answers of the attempts (correct or incorrect) on each question into consideration and detailed processes within a particular question are not included, making them insufficient to guide the question designs~\cite{long2018exactly}.
Some recent studies try to look into the inner loop (i.e., the detailed process of solving a multi-step problem), as the fine-grained interaction data is becoming available.
For example, Chris~\emph{et at.}~\cite{piech2012modeling} collected the snapshots capturing students' practice process on the programming exercises and used Hidden Markov Chain to find their common steps. However, the results are difficult for educators and question designers to interpret. Meanwhile, some studies have utilized visualizations to facilitate the interpretation of students' problem-solving behaviors. For example, Wang~\emph{et al.}~\cite{wang2017pathviewer} used a flow diagram to show intermediate results on testing cases in a programming exercise and discovered unexpected patterns caused by the unreasonable testing cases. Andersen~\emph{et al.} applied directed node diagrams~\cite{andersen2010gameplay} to emphasize certain patterns (e.g., loops) in the educational game. 
However, 
% these methods have not modeled the problem-solving behaviors in a way that can reflect students' thinking logic (i.e., the order to pass the testing cases) and engagement (i.e., the time spent and interactions such as drags-and-drops involved). Also, 
they do not support analytical tasks such as comparison of students' problem-solving behaviors from different groups. These issues hinder question designers from understanding the difficulties students may encounter and limit the practical insights distilled to guide further question designs.

%how students master the knowledge by solving a series of problem based on

% 4 Your technique + challenges & difficulty
% It is challenging to apply visual analytics to facilitate problem-solving behaviors analysis in multiple-step questions. First, how to model the massive possible answers students produce during the process and considering context information (i.e., the order to fulfill different conditions, the time spent, etc.) is challenging. Second, with each student's problem-solving process being a high-dimensional event sequence, summarizing and demonstrating their common patterns intuitively (e.g., where they get confused) is difficult. Third, comparing the similarities and dissimilarities among different groups of learners may be even harder to achieve.

% Interpret high dimensional + temporal (problem-solving logic, engagement, difficulties)
% Summarize a group of problem-solving sequences
% Compare students from different groups
% mainly contains four modules. First, it collects mouse movement data when students solve the interactive math questions on an educational platform. Second, it models the problem-solving state
%results
% 4 Your technique + challenges & difficulty
In this paper, we propose a novel visual analytics system, \textit{QLens}, to help question designers analyze students' problem-solving behaviors in multi-step questions.
The system first models students' problem-solving processes as a hybrid state transition
% from three levels 
to reflect students' problem-solving logic, engagement level, and encountered difficulties. 
Based on the modeling results, multiple coordinated views are designed to facilitate analytical tasks including interpretation, summarization, and comparison of multiple answer construction sequences at three different levels.
% at different scales. 
(1) Macro-level: the \emph{Overview} shows the overall performance achieved by a given pool of students.
% and ranks the wrong answers submitted by students by descending order of frequency.
(2) Meso-level: the \emph{Transition View} visualizes the problem-solving processes intuitively to reflect how a selected group of students proceed over time.
% In particular, a novel glyph-embedded Sankey diagram shows the approaches students take (\textit{problem-solving logic}), the amount of efforts they make to solve the problem (\textit{engagement}), and the difficulties students encountered in their problem-solving processes.
In addition, the \emph{Comparison View} enables users to compare different clusters of students in terms of these three aspects. 
(3) Micro-level: typical incorrect solution paths and the corresponding recommended paths derived from peer data, if any, are demonstrated in \emph{Transition View}
% \yong{Meng, pls fill it.}
for question designers to evaluate the feasibility of generating high-quality data-driven feedback for students in need.
Moreover, \textit{QLens} enables
% a rich set of 
rich interactions (e.g., the tooltip and filter) to show the detailed information such as common intermediate answers and facilitates the detailed exploration and inspection.
% 5 results
Three case studies with real-world datasets and detailed interviews with three domain experts demonstrate the usefulness and effectiveness of our system.

% 6 contribution
The contributions of this paper are summarized as follows:
\begin{compactitem}
\item \textbf{Interactive System:} An interactive visual analytics system,~\textit{QLens}, to help educators and question designers evaluate question designs by analyzing and comparing students' problem-solving behaviors from three levels of details. 
% understand students' detailed problem solving process, find the common struck steps, and explore the differences among groups of students. 
\item \textbf{Visualization Design:} A novel glyph-embedded Sankey diagram to represent problem-solving behaviors with the problem-solving logic, engagement level, and difficulties encountered in an informative and intuitive manner.
%Three scenarios demonstrating the usefulness of \textit{SeqDynamics} on a real-world dataset and five expert interviews showing that \textit{SeqDynamics} enhances their evaluation processes.
% \item A novel way to model problem-solving using hierarchical state transition that fits visual analytics.

% \item \textbf{Evaluations:} Three cases demonstrating the usefulness of~\textit{QLens} on real-world datasets and interviews with three experts showing that~\textit{QLens} provides sufficient information at different levels to investigate students' problem-solving processes and guide future question designs.

\item \textbf{Evaluations:}
Three case studies and interviews with domain experts provide support for the usefulness and effectiveness of~\textit{QLens} in enabling interactive investigation of students' problem-solving processes and guiding future question designs.
% \yongnote{
% Three cases
% demonstrating the usefulness of~\textit{QLens} on real-world datasets and interviews with three experts showing that~\textit{QLens} provides sufficient information at different levels to investigate students' problem-solving processes and guide future question designs.
% }{Three case studies and interviews with domain experts provide support for the usefulness and effectiveness of~\textit{QLens} in enabling interactively investigation of students' problem-solving processes and guiding future question designs.}

\end{compactitem}

%% file: sections/02-relatedwork.tex
\section{Related Work}
% This section reviews prior related work of this paper, including
The related work of this paper includes
% \yongnote{problem-solving data analysis}
problem-solving behavior modeling, problem-solving process, and event sequence visualization.

% \yong{I would suggest changing the title of Sections 2.1 and 2.2. They look too similar to each other. Also, visualization is also for data analysis.}

% \subsection{Problem-solving Process Analysis}

\subsection{Problem-solving Behavior Modeling}
% \yong{I  still do not understand the  meaning of ``outer loop'' and ``inner loop''.}
% Past analyses of problem-solving behavior mainly focus on the outer loop 
Much research has been conducted on modeling students' problem-solving behaviors. They mainly focus on the outer loop, that is how students master the knowledge by solving a series of problems~\cite{liu2017going, vanlehn2006behavior, li2020peer}. For example, Bayesian knowledge tracing was proposed to build procedural models for problem-solving processes~\cite{corbett1994knowledge}. It takes binary variables to model learners' latent knowledge. Each variable represents the understanding/non-understanding of a concept. Since learning concepts are not independent as assumed in Beyesian knowledge tracing, Learning Factor Analysis~\cite{cen2006learning} and Performance Factor Analysis~\cite{pavlik2009performance} then modeled learners' knowledge states using logistic regression with more learner features. Further, GNN has been applied to model students' performance based peer's data~\cite{li2020peer}. However, these methods only considered final answers of the attempts (correct or incorrect), which is insufficient to guide the question design.
%Performance Factor Analysis~\cite{pavlik2009performance} further considered the time it takes to learn a concept

%a Hidden Markov Model to discover "sink states", or states that the students had high probability of remaining in for several code updates.
An increasing amount of research has been done to analyze the inner loop, that is, detailed process for solving a multi-step problem, which is more complicated. Some works use machine learning methods to cluster students' detailed problem-solving behavior. For example, Chris~\emph{et al.}~\cite{piech2012modeling} recorded the snapshots of students' code during the programming, and modeled problem-solving process using Hidden Markov Model. They discovered the ``sink states'' where students cannot succeed in solving the problem once entered. In addition, they clustered all sequences to find the common patterns when students solve the programming exercise. However, they had not considered contextual information (e.g., the order how students fulfill different conditions, the time spent in each step), which are essential for understanding the reasons behind students' confusions. The states in these methods are also difficult to interpret, which hinders further analysis.
% As students' problem-solving behaviors are complex processes involving multiple stages -- problem decomposition, abstraction, and execution~\cite{stahl1994using, yadav2016computational}, only common patterns without detailed information on how students pass through those states, hinders the finding of valuable insights on how and why they get confused. 
%an overview of all the students' problem-solving behaviors (e.g., how many students transit form one state to another) and
% It is critical to figure out different stages/states to better understand how individuals reach the final answer submitted to provide personalized instruction timely and accurately~\cite{liu2017going}.
%\xm{find another accurate reference for this} 

\subsection{Problem-solving Process Visualization}
% More researchers are tackling the problem by visualizing students' problem-solving data.
Visualization techniques have also been widely applied to analyzing students' problem-solving processes.
Some studies visualized students' learning processes on solving a set of questions. For example, Xia~\emph{et al.}~\cite{xia2019peerlens} used a zip-like visualization to facilitate students to plan their personal learning path. 
Other studies applied visual analytics to students' problem-solving sequences for questions with multiple steps. Hosseini~\emph{et al.}~\cite{hosseini2014exploring} used scatter plots to represent the changes in programming concepts and whether these changes increased or decreased the correctness of the program. Furthermore, Berland~\emph{et al.}~\cite{berland2013using} used a node-link visualization to track students' program processes and the links between states show the portion of students who made the transition. PathViewer~\cite{wang2017pathviewer} modeled the state in programming exercise as a binary string of testing cases and used flow diagram to show the transition among different states. Glassman~\emph{et al.}~\cite{glassman2013toward, glassman2016learnersourcing} summarized the possible correct solutions made by students in math and circuit designs using the sankey diagram. Xia~\emph{et al.}~\cite{xia2019visual} designed transition graphs to show problem-solving logic based on the order they pass different regions of interest. However, these visualizations suffered considerably from the scalability issue due to the limitation of the modelling approach and had difficulty supporting other analytical tasks.
% \yong{Can our method handle these issues?}
%They labeled the edges between states with the percentage of participants who made that transition. Although this representation captured changes to the correctness of a student’s program, it did not fully capture the number of students in one particular state, nor the paths the students took to solve a problem. 

The task of visualizing pathways in students' problem-solving processes is similar to that video games,
% namely, how to visualize the pathways players take as they play a game.
where it is also necessary to view the pathways that game players take.
For instance,
% Chess evolution~\cite{lu2014chess} visualized potential outcomes after a piece is moved and indicates how much tactical advantage the player can have over the opponent in a step-wise manner. However, only two players are involved in the chess game.
some works on educational game~\cite{andersen2010gameplay, liu2011feature, wallner2012spatiotemporal}.
% The Playtracer tool used by Andersen~\emph{et al.}~\cite{andersen2010gameplay} and Liu~\emph{et al.}~\cite{liu2011feature} and the spatio temporal tool used byWallner~\emph{et al.}~\cite{wallner2012spatiotemporal} both 
% visualized players' states when they play  game using a node diagram.
They usually built a state transition model that stores the number of people who reached a particular state and the transition between states, and displayed similar states close to each other using node diagram. However, their states are too abstract for users to understand the semantic meanings.

% \subsection{Event Sequence Visualization and Comparison}
\subsection{Event Sequence Visualization}

% \yong{make sure that the logic flow in this section is also different from prior research papers.}

% Plenty of work has been done on event sequence visualization, including two major tasks: visual summarization and comparison~\cite{aigner2011visualization, bach2014review, guo2020survey}. 
\textcolor{black}{Plenty of research has been done on event sequence visualization and they mainly focus on two major tasks~\cite{aigner2011visualization, bach2014review, guo2020survey}: \textit{visual summarization} and \textit{visual comparison}. 
For visual summarization of event sequences,
early studies mainly summarize and visualize event sequences using timestamps and place them along a horizontal time axis, such as LifeLines~\cite{wang2008aligning,wongsuphasawat2012querying} and CloudLines~\cite{krstajic2011cloudlines}. 
% Shifting horizontally or vertically to align the event sequences are applied to a group of studies later,
Some more recent studies (e.g., EventFlow~\cite{monroe2013temporal} and EventPad~\cite{cappers2017exploring}) have also shifted the event sequences to make them better aligned horizontally or vertically. 
Other studies also considered using the circular or spiral layout to summarize event sequences visually. For example,
% utilize its similarity with the clock and the time bin of twenty-four hours, e.g., 
ClockMap~\cite{fischer2012clockmap} 
is designed to show 24-hour network traffic and SpiraClock~\cite{dragicevic2002spiraclock} visualizes the upcoming events with a clock-like design. 
% \yong{Meng, pls fill this.} 
To improve the scalability of event sequence visualization, hierarchical visualization is also proposed to further summarize event sequences, for instance, Timeline Trees~\cite{burch2008timeline} and Sequence Surveyor~\cite{albers2011sequence}. Further, advanced data mining and machine learning methods are utilized to summarize the common patterns~\cite{kultys2014sequence, sakai2014sequence, chen2017sequence, guo2018visual, guo2017eventthread}.
% Sequence Bundles~\cite{kultys2014sequence} and Sequence Diversity Diagram~\cite{sakai2014sequence} plot the event types to Y-axis and then use the edge bundling to summarize the event type transition overtime.
% For example, Sequence Synopsis~\cite{chen2017sequence} applied the MDL (minimal distance length) to calculate the representative sequence of a collection of sequences. Eventhread~\cite{guo2018visual, guo2017eventthread} applied an unsupervised stage analysis algorithm to identify the semantical stage in event sequence data.}
% \yong{The overall logic of this paragraph is not quite clear. I do not quite understand it.}
Event sequence comparison mainly contains three types of comparisons among sequences, including one-to-one sequence comparison, one-to-many sequence comparison, and comparison of two collections of sequences.
For example, Similan~\cite{wongsuphasawat2009finding} compares two sequences by separately comparing each individual event.
% based on individual events.
Matrixwave~\cite{zhao2015matrixwave} uses a specialized matrix design to compare two sequences of clickstreams. As for the one to many sequences, EventAction~\cite{du2016eventaction} shows the sequences by ordering multiple similar sequences according to their similarity with the target sequence to facilitate easy comparison. COCO~\cite{malik2015cohort} and integrate statistical information and the evolution of a group of sequences.}

% \xm{
\textcolor{black}{However, the visual analysis of problem-solving process in our scenario is more complicated than the visual analysis of sequences in prior studies.
% those in the previous works. 
First, we need to compare multiple groups of sequences rather than only summarizing one group. 
% We follow COCO~\cite{malik2015cohort} and also integrate statistical information of visual analytics to facilitate the comparison of two groups of sequences. \yong{pls fill the whole and check if it is correct.}
Second, more than one event can appear at the same time, as students can fulfill different conditions at once. 
% Such kind of event sequences  was not able to be processed by COCO. 
Thus, new visual designs need to be proposed to enable analytical tasks of problem-solving behaviors on multi-step questions.}

%% file: sections/03-system.tex
\section{System Overview}
\label{sec-system-overview}

Our system was designed on the basis of real-world requirements of question designers to analyze and determine students' problem-solving behaviors to better design the questions. In our application scenario, we collaborate with one educational company that offers interactive math questions online for 30+ elementary and middle schools and tens of thousands of students.
%They also design questions for the World Class Tests(WCT), an international math problem-solving test. 
The online learning platform they run contains 1718 mathematical questions. We interviewed four domain experts (E1 - E4) from the company. E1 is the general manager; E2 is the product manager; E3 and E4 are the curriculum and question designers. We list the following four primary design requirements (R1-R4) derived from the interviews that guided the system design.

\textbf{R1: Show students' overall problem-solving performance.}
Experts need an overview about the problem-solving results from students towards the questions they designed. For a particular question, they want to know
% the number of students that have tried this question, 
what grades are the students from, what are the scores they get and the time they spent. In addition, E3 mentioned that the common incorrect answers can be summarized and ranked to help question designers grasp a general understanding of students' knowledge level and the problem difficulty.

\textbf{R2: Summarize and present the multi-step problem-solving behaviors.}
Experts require a summarization and presentation of the multi-step problem-solving behaviors to reflect students' logic of thinking, engagement, and difficulties encountered. E3 mentioned that the questions they designed aim to cultivate students' computational thinking and encourage creative ways to interact with the question. Nearly all the questions have the knowledge they hope students to grasp. Thus, they required the system to summarize students' approaches for them to understand whether students' problem-solving processes are consistent with the intent of the question they designed and how much effort is needed to solve the problem. E3 and E4 added that, they also want to identify where and when students get stuck, which would be beneficial for designing feedback to students. 
% (e.g., post-question solution or hints).

% abstract final performance and granularity

% temporal semantics

% All experts mentioned that the system should support level-of-detail analysis of students problem-solving behaviors on each problem. From the overall score distribution, common incorrect answers to the step-wise pattern of a group student, and the detailed trajectory of a particular student. Especially, E3 and E4 suggested that the system should design visualizations that demonstrate students' problem-solving logic as it is very crucial for them to evaluate the question design in terms of the testing points setting. E1 also mentioned that they need to know how much efforts (e.g., time) students put on interacting with these questions. They also care about the confusion (where and when students get struck) for the feedback design (e.g., hints or solution) of the question.
% State of confusion (where and when students get struck) among the students in a particular problem: The system should incorporate visuals that highlights the area of confusion as it is very crucial for the instructors to understand where most students get confused or find it difficult to surpass a particular stage in a problem. 
 
\textbf{R3: Enable the comparison of students from different groups.}
The system should also enable question designers to compare two or more groups of students with different grades or scores and explore their differences in the logic of thinking and engagement. As mentioned by E1, each problem has its main targeted group of students (e.g., a certain grade) even though they allow students from different grades to access the question. E3 and E4 said they hope the question can be assigned to a group of students who can interact more with it while not feeling too difficult, which is based on the zone of proximal development (ZPD) principle in education~\cite{veresov2004zone}. Thus, he wanted to know whether students from different grades confront different difficulties and which groups the question is more suitable for.
% In addition, highlighting and updating users’ selections simultaneously in multiple coordinated views to achieve consistency are also needed. 

\textbf{R4: Evaluate the feasibility of providing data-driven feedback.}
E2 mentioned that they consider enhancing the current platform by providing data-driven feedback, e.g., recommended solution paths or next-step hints using existing data. For example, providing recommendations based on correct paths produced by existing users to those students who tried the incorrect answers. However, they are unsure whether this method can be applied based on the existing data. He suggested that the system provides a way for them to evaluate the feasibility of such data-driven approaches. This could also give them insights into how to design on-the-fly guidance in the future. 

\begin{figure}[t!]
  \centering 
  \includegraphics[width=1.0\linewidth]{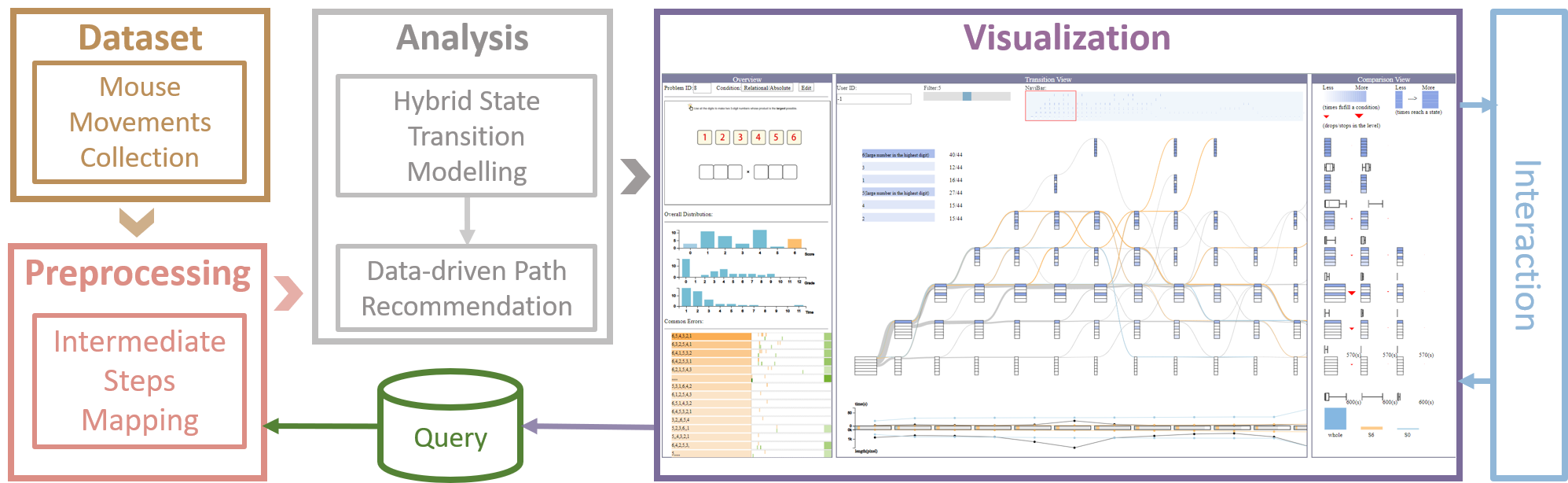}
  \vspace{-0.7cm}
  \caption{System overview. It contains preprocessing, analysis, visualization, and interaction modules.
%   It contains four modules: data preprocessing, analysis, visualization, and interaction.
}
   \label{fig:sys-overview}
\end{figure}
% \textcolor{red}
Based on the aforementioned requirements, we have designed \textit{QLens} to provide a visual representation and an enhanced analysis of problem-solving processes and behaviors for question designs. Fig.~\ref{fig:sys-overview} displays the overview of the system architecture, containing four modules: (1) data collection module collects mouse movement data
% during the problem-solving process
and preprocesses it to sequences of intermediate answers indexed by learners' IDs (R1); (2) analysis module models the problem-solving steps as hybrid state transitions (R2, R3) and implements a data-driven solution recommendation system (R4); (3) visualization module makes use of linked views to facilitate interpretation and comparison of problem-solving behaviors
% of different groups in context 
(R1, R2, R3, R4); and (4) interaction module supports exploration based on users’ preference or input (R3, R4).

%% file: sections/04-model.tex
% \section{Problem Characterization}
\section{Data and Modeling}
% \yong{I have changed the title. Pls confirm if it is OK.}
In this section, we introduce the mouse movement data and its preprocessing, the problem-solving behavior modelling, and the data-driven feedback construction.

\subsection{Data Preprocessing}
We have collaborated with an online education company and collected students' problem-solving data from their popular online education platform, which offers interactive math questions/quizzes online to over five thousands students from elementary and middle schools. 
% \yong{Meng, pls fill the number.}
% from 30 elementary and middle schools.
% with tens of thousands of students. 
We focus on problem-solving records from April 2019 to January 2020, which consist of 2,30,644 records from 5,266 students and 1,718 mathematical questions. 
Each record consists of a question ID, a student ID, and the mouse movement data along the problem-solving process. 
The mouse movement data contain mouse positions, mouse event types (i.e., up, down, and move), and time stamps. Specifically, "move" event is recorded at most 50 times per second; "up" and "down" are recorded whenever they are triggered. Students' grade and question content information are also recorded and further considered in our analysis.
% \by{(Do we need to mentioned that we only analyze 8 questions?)}

\textcolor{black}{To understand how students solve a multi-step question, we have to map the raw mouse movement trajectories into a sequence of steps. 
% Here, a step is defined as ``\textit{the smallest user interface interaction for which it makes sense to call it correct or incorrect}''~\cite{vanlehn2016regulative}. 
More specifically, for our scenario of analyzing multi-step problem-solving behaviors, a step is the smallest user interface interaction that changes the intermediate answer (e.g., $6, null, null, null, null, null$) filled in the blanks. The mapping from raw mouse movement trajectories to steps can be generalized to other multi-step questions according to the concrete format of the answer.}  

\begin{figure}[t!]
  \centering 
  \includegraphics[width=1.0\linewidth]{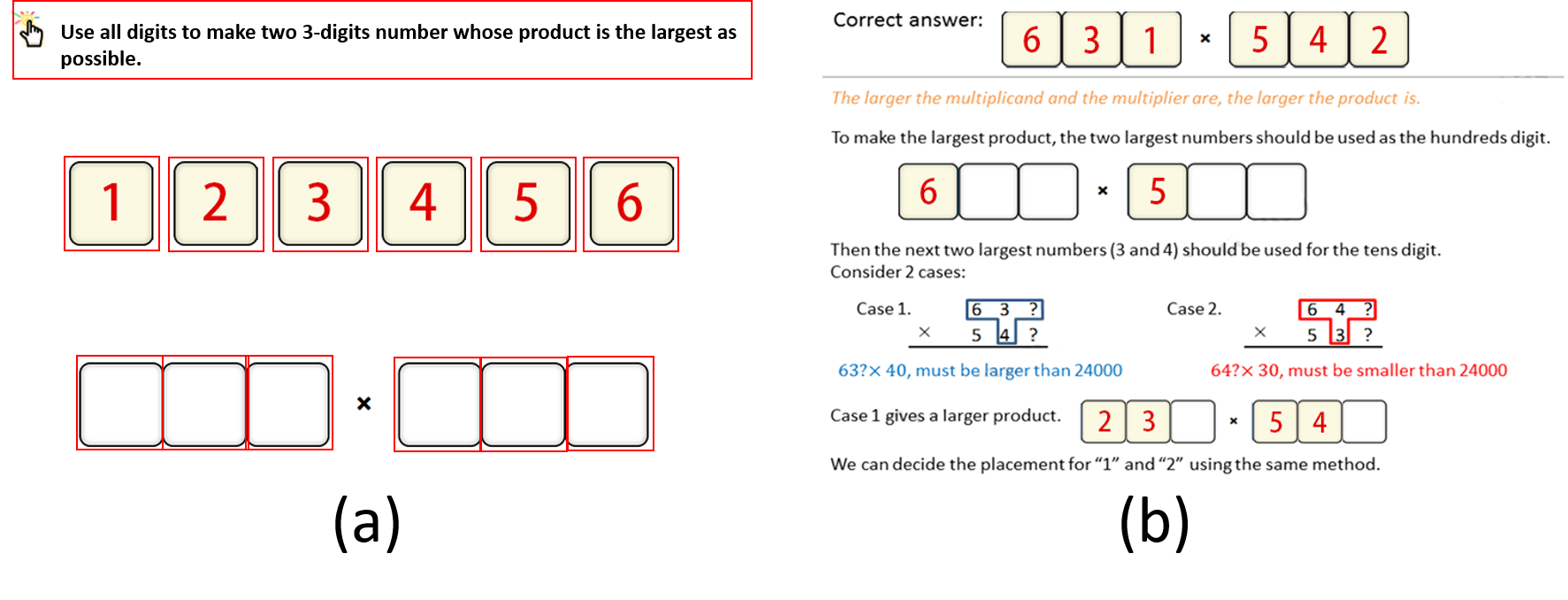}
  \caption{An example of interactive math question (a) and its solution (b).}
   \label{fig:example}
\end{figure}

\textcolor{black}{For each question, we first extract the regions of interest (ROI) by drawing bounding boxes around all graphical components using Canny's edge detection algorithm~\cite{green2002canny},
% in Open-CV library, 
as shown in Fig.~\ref{fig:example}(a). We then remove undesirable bounding boxes that are not interactive using predefined rules (e.g., the question description area). Then, we number the ROIs as 1,2,3, etc., with an order from the top left to the bottom right. 
For each student, we generate an ROI sequence by checking whether each mouse position lies inside one of the ROIs and replacing it with the ROI number and 0 if the mouse position does not fall within ROIs. 
% \yong{what do you mean by the prior sentence?}
Since only drag-and-drop mouse interactions (a mouse down followed by a mouse up event) ~\cite{wei2020predicting} may change the answers, we make ROI pairs if two consecutive ROIs have mouse down and mouse up events, respectively.
We then construct the paired ROI sequence 
% based on the original ROI sequence. From the paired ROI sequence, we
and generate an intermediate answer sequence. 
For example, initially the intermediate answer sequence will be all null, i.e., {null, null, null, null, null}. After the first interaction, we replace the null value with values from the corresponding ROIs and this is repeated for all the interactions throughout the whole problem-solving session. At the end of this process, we arrive at the final answer of the student for that particular question.}
% To understand how students solve a question, we first map the raw mouse movement trajectories into a sequence of intermediate answers, which the students filled along the way in solving a question. We detect the bounding boxes of the graphical components in a question using OpenCV library, as shown in Fig.~\ref{fig:example}(a). 
% We then delete component(s) that cannot be interacted with drag or drop using pre-defined rules (e.g., the question description area). 
% % \yong{manually? What do you mean by saying manually?}
% Finally, we recover the intermediate answers students fill in those blanks during the problem-solving process based on mouse events (i.e., up and down) and the interactive areas.

\subsection{Problem-solving Behavior Modeling}
% As for \textbf{R2} and \textbf{R3},

To reflect the problem-solving logic, engagement level, and difficulties students encountered and facilitate further visualization tasks,
% in the representation, 
we use a hybrid state Markov Chain to model students' problem-solving behaviors~\cite{abate2008markov}. Before introducing the detailed problem-solving behavior modeling, we first define the following terms as below:

% The related items are defined as follows:
\begin{compactitem}
\item \textbf{Condition:} One criteria that students need to fulfill to get part of the score.
% \yong{Can the definition be more specific? The current definition is vague.}
For example, in Fig~\ref{fig:example}(a), placing one of the six digits in the correct position accounts for one condition being satisfied.
% whether one of the six digits is put to the right position is one condition.

\item \textbf{Condition array:} a string consisting of $0$ and $1$ to indicate whether a set of conditions are fulfilled or not. 
The length of the string is the number of the conditions. For example, the length is six for the question in Fig~\ref{fig:example}(a).

\item \textbf{Stage:} the number of conditions the current answer fulfills. For example, for the question in Fig~\ref{fig:example}(a), $(6, null, null, null, null, null)$ fulfills one condition and at Stage one.

\item \textbf{Time elapse:} the time period between the time a student starts to solve the problem and the time he/she reaches a certain step.

\item \textbf{Trajectory length:} the length (pixels) of the cursor moves on the screen from the time the student starts to solve the problem to the time he/she reaches a certain step.
\end{compactitem}

Based on the definitions above, \textbf{State ($S$)} is modeled as a two-level hybrid structure:
\begin{compactitem}
\item \textbf{Level1: \{Step, Stage\} + \{Condition array, Time elapse, Trajectory length\}}
\item \textbf{Level2: \{Intermediate answer\}}
\end{compactitem}

%\yong{Do not quite understand the above signs.}

For a particular student, the change of the conditions satisfied along the way can reflect the problem-solving logic. If a student's stage cannot steadily go up as the step continues (e.g., the stage remains the same for several steps or drops continuously),
% this can be interpreted as 
it often indicates that a student probably
encounters difficulties, as suggested by the domain experts. For example, if a student's problem-solving sequence is as follows: $(6, null, null, null, null, null)$, $(6, null, null, 5, null, null)$, $(6, 4, null, 5, null, null)$, $(6, 4, 3, 5, null, null)$, $(6, 4, 3, 5, 2, null)$, $(6, 4, 3, 5, 2, 1)$, this shows that from the third step, he/she cannot fulfill more conditions and stays at Stage two (the correct answer is $(6, 3, 1, 5, 4, 2)$). 
% One is the key part \{$ST$, $LE$\} and another one is the contextual part \{$CO$, $TE$, $TL$\}.
% \item \textbf{Problem-solving sequence} Each problem-solving sequence is an ordered sequence of states \{$S_0, S_1 ..., S_i$\}, where $S_i = \{ST_i, LE_i, CO_i, TE_i, TL_i\}$.

We introduce the two-level hybrid state because of the following reasons. First, \{$Step$, $Stage$\} in \emph{Level1} keeps the number of possible states in a reasonable size. 
If we directly use the answers student fill in as the state, then a huge state space would appear. As shown in Fig.~\ref{fig:example}, given six blanks and six possible answers, there would be 5040 ($7*6*5*4*3*2$) possible outcomes \textcolor{black}{(starting from 7 instead of 6 as students can fill nothing in each blank)}. 
With such a big search space, it is difficult to 
% present and mine 
extract and further visualize
the transition patterns among these states.
% using visualization. 
Previous work~\cite{wang2017pathviewer} used one binary bit to represent one testing case in the programming exercise and a string of binary bits (0 or 1) to represent the state, which can reduce the state space. However, the state space is still large. If we consider one testing case as one number being correctly placed in one blank, then the state space is 64 ($2^6$) for the given example. Inspired by a previous study~\cite{xia2019peerlens}, we use the number of conditions to be satisfied as the key component of the state, which reduces the state space to $7$ in the given example. 
By doing this, it is intuitive to show the $Step$ and $Level$, as it captures how a learner proceeds along the way, i.e., whether he/she is progressing towards the correct solution or deviating from it.
However, only representing the state using step and stage is still too coarse.
Further information should be added to provide additional contextual and semantic meanings for diagnosing where and why students get confused, e.g, spending more time to fulfill a certain condition. Thus, \emph{Level 1} \{Condition Set, Time elapse, Trajectory length\} enriches the state with contextual information.

However, a problem arises if we only consider \emph{Level 1}. One condition set may contain many possible intermediate answers. For example, $(6, 3, null, null, null, null)$ and $(6, null, null, 5, null, null)$ satisfy two conditions while they represent different thinking styles. Without the detailed answers, recommending the path or next step for students who currently have not satisfied all the conditions is difficult. 
Thus, we represent the intermediate answer in \emph{Level 2}. Such information can be interactively shown in the visualization during the exploration of a user.
% when required by the user.
%According to the state, we then construct the hybrid state Markov Chain of a group of learners' problem-solving behaviors from two levels. 

By using the states, we finally construct a hybrid State Markov Chain to model the problem-solving behaviors of a group of students from two levels.
\emph{Level 1} is mainly used for visualization and \emph{Level 2} is mainly used for data-driven path recommendation. In summary, 
% by constructing the state into two-level hybrid state transition, 
we can display and utilize students' problem-solving information for different tasks intuitively and effectively by constructing the states into two-level hybrid state transition.

\subsection{Data-driven Feedback Construction}
\textcolor{black}{According to the requirement (R4) raised by domain experts, question designers need to evaluate the feasibility of using existing data to provide data-driven feedback to guide students' problem-solving process.
% For example, which incorrect answers are the most common one and less student can pass; which conditions partial students can fulfill and whether the sequences from those partial students can be used for the feedback design. 
The common questions that question designers want to explore
include: Which incorrect answers are the most common one that fewer students can pass? Which conditions can students fulfill? Can the sequences from those students be used for the feedback design?
Since question designers focus on the feasibility instead of the quality of the feedback design at the current step, we selected a basic path recommendation algorithm adapted from Markov Chain model to show the representative answers students fulfill~\cite{xia2019peerlens}, where the next step is chosen based on the highest transition probability.
% As for \textbf{R4}, 
% \yong{What is the logic connection between this part and the previous part? Why data-driven feedback construction? Need to clarify them here.}
% When constructing data-driven feedback, the basic idea is to recommend the paths conducted by students who can correctly solve the question to students who cannot.
% % according to experts' suggestion. 
% We implemented a popular path recommendation algorithm based on the path similarity applied in previous research~\cite{du2016eventaction, xia2019peerlens}. 
% To be specific, 
% More specifically,
% for a student who submitted an incorrect answer, we find all the problem-solving paths (1) that fulfill all the conditions at the end and (2) whose states cover the final states resulting the incorrect answer.
More specifically,
for an incorrect answer, we find all the problem-solving paths (1) that satisfy all the conditions at the end and (2) whose states cover the incorrect answer.
% \yong{Meng, pls check if I change your idea.}
% passed through the incorrect answer.
If more than one path satisfies these constraints, then we construct the data-driven path according to the transition probabilities among different states and select states with the highest transition probability.}
% If more than one path satisfies this constraint, then we rank them according to their similarities with the current path. The similarity is calculated by the total steps, the total time, and the total trajectory length. Finally, we recommend the solution path with the highest similarity. 
% If we cannot find learners who satisfy the two constraints, we consider the penultimate state of the learner in the same until find such path(s).
% model a group of problem-solving sequences as a Markov Chain (MC) based on Level1. 

% step: every time when the answer is changed, it's called a step

% level: how many conditions are satisfied

% state: (step, level, time elapse, trajectory length, conditions)

% transition: from one state to another state

%% file: sections/05-visualization.tex
\section{Visualization}
\subsection{Design Tasks}
Based on requirements (R1-R4) from expert interviews and the problem-solving model, we have derived the follwoing design tasks (T1-T6).
% \textbf{R1: Show the overall problem-solving results from students.}
% \textbf{R2: Summarize and Present the multi-step problem-solving behaviors to reflect students' logic of thinking and engagement.}
% \textbf{R3: Enable the comparison of different student groups from both the logic of thinking and engagement.}
% \textbf{R4: Evaluate the feasibility of constructing data-driven feedback.}

\textbf{T1} Show the overall performance of students' problem-solving behavior~\textbf{(R1)}.
The visual design need to provide the overall distribution of students' scores, grades, and time spent to facilitate question designers to evaluate the overall performance and the difficulty level of the question at a glance. Common incorrect answers should also be ranked according to the frequency for question designers to check students' understanding.
% \textbf{T2} Display the common errors students have when solving the problem.

\textbf{T2} Demonstrate students' problem-solving logic intuitively~\textbf{(R2)}.
The visualization should demonstrate students' problem-solving logic by showing the order they fulfill different conditions step by step. In addition, it should show intuitively where students get confused and what are the common approaches taken, for questions designers to better understand the problem-solving behaviors of a group of students.

\textbf{T3} Display the efforts (time and trajectory length) students pay along the way of problem-solving~\textbf{(R2)}.
Question designers want to know whether students take the question seriously and how much effort they pay on the question. 
Thus, the time they spent and the trajectory length of the cursor should be displayed to show whether students are engaged in the problem-solving process.

\textbf{T4} Support the comparison of the problem-solving logic, engagement, and difficulties of different groups from multiple granularities~\textbf{R3}.
The visualization designs should enable the comparison of problem-solving behaviors (i.e., the problem-solving logic, engagement, and difficulties) from students with different grades or scores. The design should support the comparison at both a high level (i.e., statistical information such as the total time spent) and a detailed level (i.e., step-wise information such as condition fulfilled on each step).

% \textbf{T5} Compare problem-solving behaviors from different groups in detail.
\textbf{T5} Facilitate the evaluation of data availability for providing data-driven feedback~\textbf{(R4)}.
To improve the question design by providing data-driven feedback on students who cannot solve the question, question designers need to evaluate whether the existing data covers the common errors. The visual design should show the common errors and the number of students that bypassed these errors and solved the question correctly at the end for the availability evaluation.
% -> confidence

% Provide the confidence for the data-driven solution recommendation.

\textbf{T6} Demo the data-driven feedback to verify the confidence for providing such feedback~\textbf{(R4)}.
For the data-driven feedback for each incorrect answer, question designers need to check the detail of how many students make that error and what is the data-driven feedback, to verify the confidence for providing such feedback.
% Enable the verification of the confidence for providing data-driven feto Display the typical erroneous path and the data-driven recommendation path.

\subsection{Visual Design}
We present a novel visual analytics system~\textit{QLens} to accomplish the aforementioned tasks. This system can aid educators and question designers in exploring, analyzing, and understanding the problem-solving processes of students in multi-step questions. The visual analysis module of \textit{QLens} incorporates the following three views. 1) \textbf{Overview}, as shown in Fig.~\ref{fig:teaser}(a), displays basic information about the students, which includes the background of students, score distribution, time spent by students and also information about common errors (T1, T2, T6). 2) \textbf{Transition View}, as shown in Fig.~\ref{fig:teaser}(b), visualizes the steps and stages involved in solving a problem which helps in understanding the different approaches and problem-solving logic that the students apply to solve the problem. This view also displays the data-driven recommended paths of common errors for question designers to evaluate the feasibility for providing data-driven feedback (T3,T7). 3) \textbf{Comparison View}, as shown in Fig.~\ref{fig:teaser}(c), facilitates a detailed comparison of the problem-solving process of two or more groups of students based on users' selection (T4, T5). Moreover, a collection of interactions, such as filtering, highlighting, and tooltips, is also available for the users to explore the dataset freely.
\begin{figure}[t!]
  \centering 
  \includegraphics[width=1.0\linewidth]{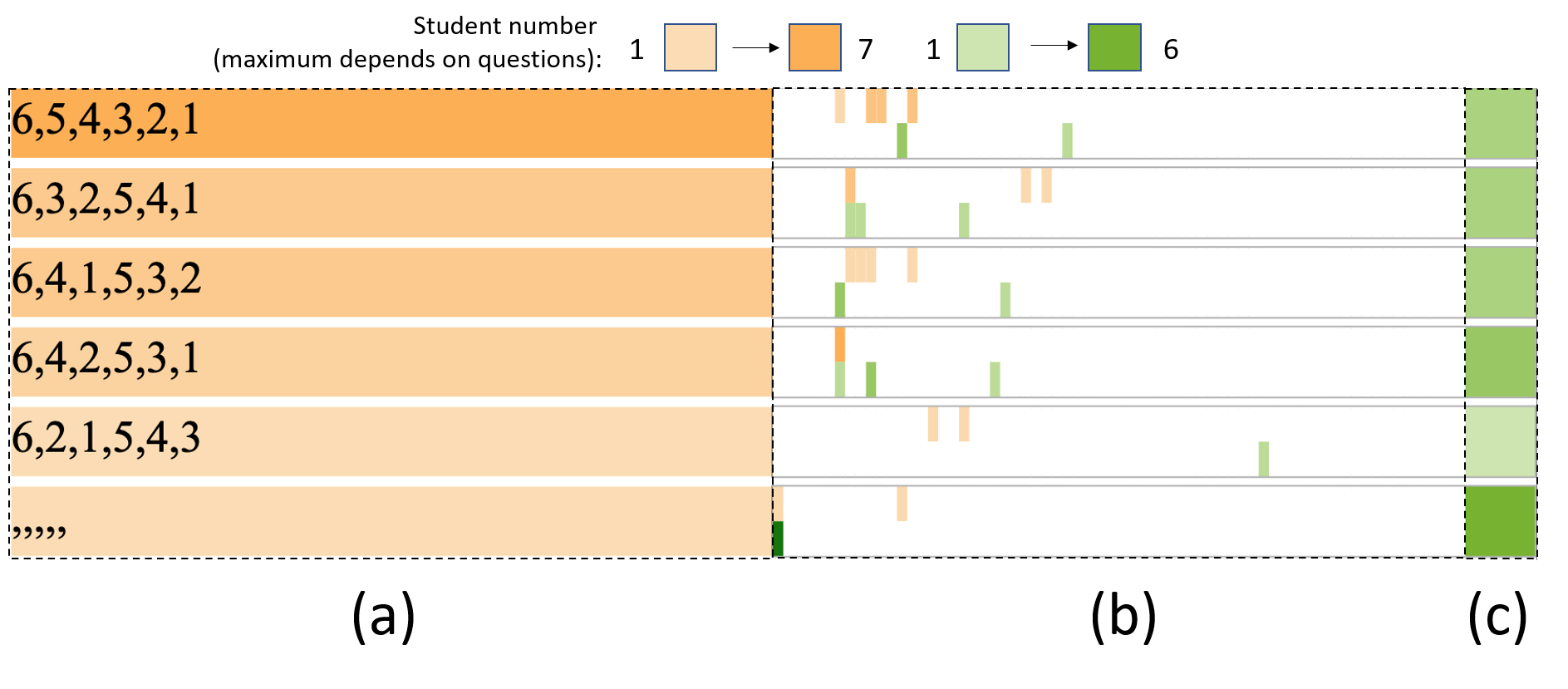}
  \vspace{-0.4cm}
  \caption{The common error panel. The ranked list of common errors based on descending order of frequency (a), the step distribution where common errors appear in both incorrect paths (orange) and correct paths (green) (b), and the number distribution of how many correct paths passing each common error (c).}
   \label{fig:common_errors}
\end{figure}

\begin{figure}[t!]
  \centering 
  \includegraphics[width=0.8\linewidth]{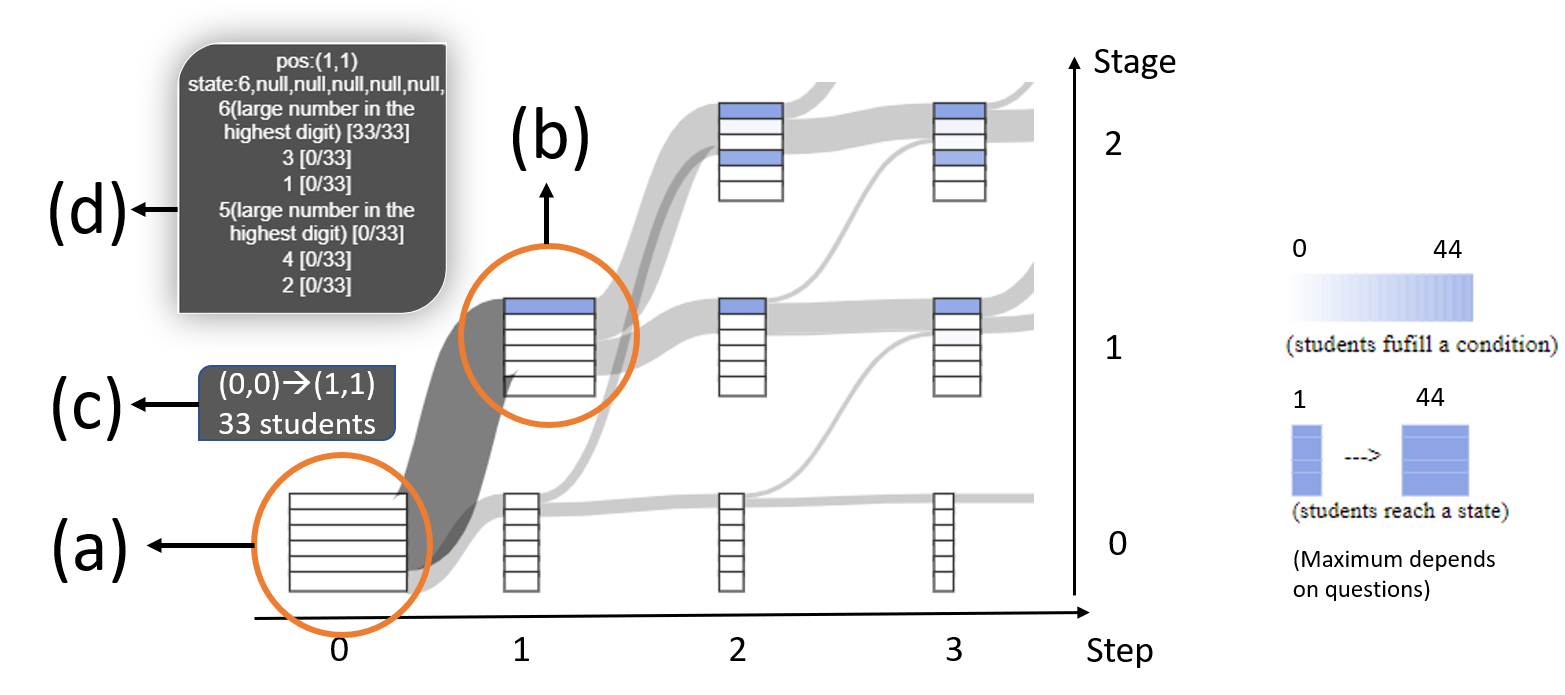}
  \vspace{-0.4cm}
  \caption{The enlarged part of the transition graph in Fig.~\ref{fig:teaser}(b2).}
   \label{fig:transtion}
\end{figure}

\begin{figure}[t!]
  \centering 
  \includegraphics[width=1.0\linewidth]{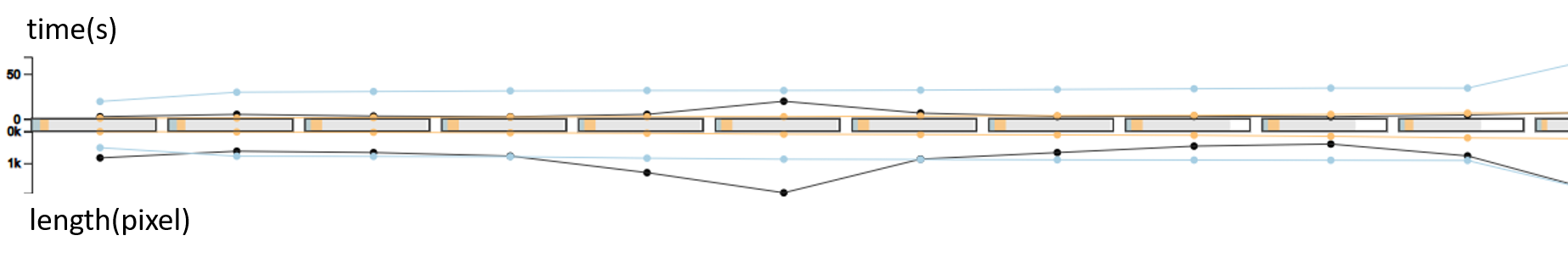}
  \vspace{-0.4cm}
  \caption{The contextual line chart with dual axes. The upper line chart shows the average time spent (seconds) on each step and the lower line chart shows the trajectory length of the cursor (pixels) on each step. Lines of different colors represent different groups.}
   \label{fig:line_chart}
\end{figure}

\subsubsection{Overview}
The Overview (Fig.~\ref{fig:teaser}(a)) aims to provide a macro-level view of the information about the students and their overall performance using distribution charts and zipper-like visual metaphor. This view aids users to acquire a comprehensive understanding of students' knowledge level, their background, engagement in the problem-solving process, and the difficulty level of the question. 

The Overview has three panels. 
%in terms of scores, time spent/efforts made, and common incorrect submissions made by the students 
\textbf{Question panel} (Fig.~\ref{fig:teaser}(a1)) displays the preview of the question for which the mouse trajectories are being analyzed. \textbf{Distribution panel} (Fig.~\ref{fig:teaser}(a2)) shows the distribution of students among different scores, grades, and also time invested in the problem-solving process using the bar chart. The y-axis always denotes the number of students and the x-axis denotes the scores, grades, and time (in minutes) respectively. \textbf{Common errors panel} (Fig.~\ref{fig:teaser}(a3)) gives an overall insight into the errors student made when solving the problem. Error related information is also displayed, such as the number of students who encountered the error, number of students who bypassed the error, where they encountered the error and so on. 
% and also how these information can be used to indicate whether the data is sufficient to support the data driven feedback (e.g., recommended path) provided to the students

The common error panel has three parts. Fig.~\ref{fig:common_errors} is an enlarged version of Fig.~\ref{fig:teaser}(a3). Fig.~\ref{fig:common_errors}(a) shows a ranking list of the common incorrect answers. The color encoding is based on the frequency of occurrence, where the frequent error has a darker shade of orange, and vice versa. Fig.~\ref{fig:common_errors}(b) uses a zipper-like visual metaphor to highlight the steps where the students encounter these errors. For each common error, the upper teeth uses the orange color to indicate the steps where students who cannot solve the question submit these incorrect answer. Similarly, the lower teeth uses the green color to indicate the steps where students who can solve the question encounter the same error. The zipper-like visual metaphor also uses a color gradient to encode the number of submission information. For example, the step where more number of students submit the incorrect answer has darker shade and lighter shade indicates less number of submissions. \textcolor{black}{If the error steps appear at the beginning of the zipper-like metaphor, then this error was more likely to be caused by  students’ carelessness, while later represents that the error is difficult to overcome.}
 Fig.~\ref{fig:common_errors}(c) shows the number of students who were able to overcome these common incorrect answers. Here the color gradient is used to encode the number, darker the shade higher the number of students. 

Further, this view also supports interactive options such as selecting the question of interest and filtering the group of students based on the score or the grade. It also enables question designers to edit the conditions and provides two types of conditions by default: absolute and relational.
\textcolor{black}{Relational condition utilizes the relationship between different interactive elements in the question to construct the required conditions to be fulfilled, whereas absolute condition considers the actual position of the interactive element in answer to construct the required conditions. 
% \yong{what do you mean by ``blank to construct"?}
For example, if element\_1 is greater than element\_2 then element\_1 > element\_2 can be one of the several Relational conditions to be fulfilled. Likewise, placing element\_1 in first the answer blank becomes an absolute condition.}

% Absolute conditions are constructed according to whether the elements are put to the absolute correct positions while the relational condition is considered by the relationship of the interactive elements. For example, ``1 < 2'' is a relational condition and ``1 is placed to the first position'' is a relational condition.

\subsubsection{Transition View}
The Transition View (Fig.~\ref{fig:teaser}(b)) aims to provide a holistic view of the multi-step problem-solving behavior to reflect students' problem-solving logic, engagement, and difficulties encountered by using a glyph embedded Sankey diagram and a contextual line chart. The view has three parts: control panel (Fig.~\ref{fig:teaser}(b1)), transition graph (Fig.~\ref{fig:teaser}(b2)), and engagement chart (Fig.~\ref{fig:teaser}(b3)). 
\textbf{Transition graph}, as shown in Fig.~\ref{fig:transtion}, intuitively visualizes how a group of students fulfill a set of conditions step by step in order to reach the final answer (i.e., the problem-solving logic). The x-axis represents the step and a student makes a step when he/she changes the answer by deleting/inserting/exchanging/translating element(s). The y-axis represents the stage, where the stage is the number of conditions fulfilled in each step. For example, if a student has a sequence starting with $(null, null, null, null, null, null)$, $(6, null, null, null, null, null)$, he/she moves from Fig.~\ref{fig:transtion}(a) (Step zero, Stage zero) to Fig.~\ref{fig:transtion}(b) (Step one, Stage one). By placing 6 in the correct position, one condition is fulfilled. A student moves to a higher stage only when one or more conditions are satisfied otherwise the student remains in the same stage or drops to a lower stage if he/she breaks one or more condition(s) no matter how many steps he/she makes. The transition lines between the glyph show the number of students going from one state (step, stage) to another. For example, as shown in Fig.~\ref{fig:transtion}(a), all students fulfill zero conditions at the beginning. The 33 students (Fig.~\ref{fig:transtion}(c)) move from Stage zero (zero conditions fulfilled) to Stage one (one condition fulfilled) at the second step. By hovering the mouse to the glyph, we can check the details like the number of students fulfilled each condition (Fig.~\ref{fig:transtion}(d)). \textcolor{black}{When a student is selected, his/her path will be shown in red in the transition graph and the recommended path, if any, will also be shown in green (as highlighted in Fig.~\ref{fig:teaser}(b)). The corresponding intermediate answers are shown in different lines to avoid overlap, with green text representing the recommended path.}
% only when any one of the conditions required by the question is satisfied.
% Student may also drop from when his/her subsequent steps/moves break the already fulfilled condition.  
% in terms of scores, time spent/efforts made, and common incorrect submissions made by the students by summarizing the entire problem-solving process in terms of steps and stages 
\begin{figure}[t!]
  \centering 
  \includegraphics[width=0.8\linewidth]{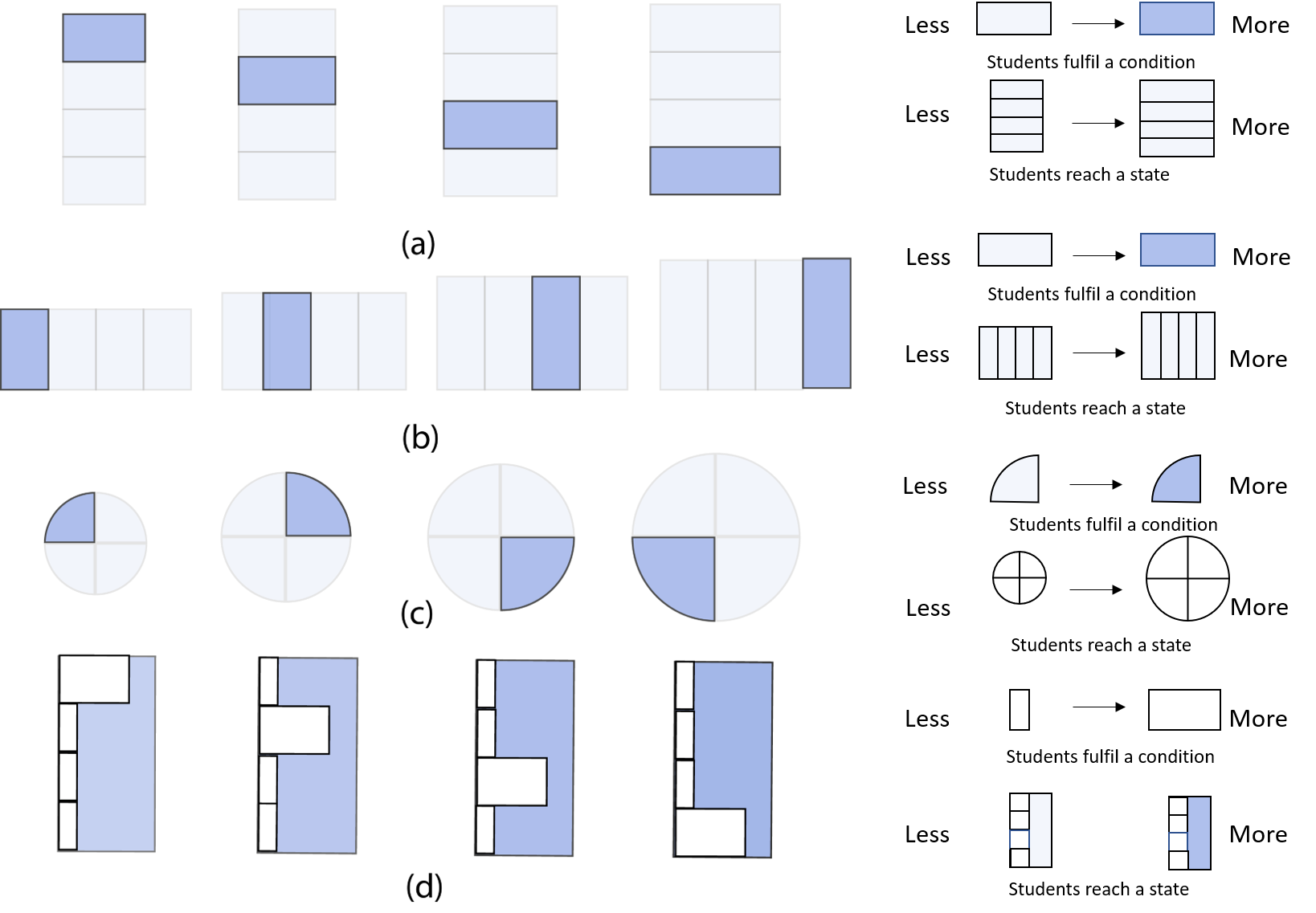}
  \vspace{-0.5cm}
  \caption{The condition glyph (a) in \textit{QLens} and three alternative designs (b) (c), and (d).}
   \label{fig:glyph}
\end{figure}

\textbf{Condition Glyph.} As shown in Fig.~\ref{fig:glyph}(a), condition glyph is designed to describe how a group of students fulfill the conditions at a certain step and stage. The glyph we designed consists of a block of vertically stacked rectangles. The number of rectangles in the block is determined by the number of conditions to be satisfied in order to solve the question, and each rectangle represents a specific condition. For example, question in Fig.~\ref{fig:teaser}(b) has six conditions(Fig. ~\ref{fig:example}). The number of students who fulfilled the conditions is encoded using color. The dark shade in the rectangle indicates more number of students have fulfilled the condition, whereas a relatively lighter shade or no shade indicates fewer students or no student have satisfied the condition. The width of the glyph represents the number of students who have reached/dropped to that state (step, stage).

\textbf{Glyph Alternatives.} The condition glyph is designed and refined several times based on feedbacks/comments from the four experts. We came up three alternative designs in the process which were later rejected for different reasons. The first alternative design (Fig.~\ref{fig:glyph}(b)) is much similar to our final design except that the rectangles are horizontally stacked. However, this design was rejected because, the number of steps is larger than the number of conditions for all questions. Less information would be shown on one page if we use (b). The second alternative design (Fig.~\ref{fig:glyph}(c)) is adapted from the paper~\cite{wu2015egoslider} to use a pie chart to show the condition distribution. Each sector represents a condition. Different from the work ~\cite{wu2015egoslider} that uses the angle to show the portion of different conditions, we use the color to show the number of students that fulfilled each condition to guarantee that the position of each condition is fixed to facilitate comparison among different glyphs. The area of the circle represents the number of students. This design was not adopted since we focus on the order of how students fulfill these conditions. For example, a large difference exists between students who start with the first condition and students who start with the last condition. As shown in Fig.~\ref{fig:glyph}(b),
such differences cannot be reflected explicitly. The alternative design Fig.~\ref{fig:glyph}(d), uses the width of the bar to show the distribution and the color of the background to present the number of students in that block. \textcolor{black}{The alternative design (d) is accurate in terms of comparing the number of students within a glyph. However, we abandoned this alternative,
% for the reason that
%
% First, the background color is partially covered by the distribution in different degrees, which may affect peoples' perception. Second, if no student fulfills two or more consecutive conditions, then figuring out the total number of conditions as well as the index of each condition would be difficult.
%
% if no student fulfills two or more consecutive conditions, then figuring out the total number of conditions as well as the index of each condition would be difficult.
since it will be difficult to figure out the total number of conditions as well as the index of each condition from this alternative design, if no student fulfills two or more consecutive conditions.
}

% ~\cite{jardine2019perceptual}.
% Combining the experts' feedback, (Fig.~\ref{fig:glyph}(a)) is more intuitive to capture the distribution and the changes between two glyphs.
% ~\by{Add more alternative designs and the user study}

% TheHere steps indicate the number of meaningful interactions (drag and drop) made by the students (i.e.) when the answer is changed it is considered as one step. 
% And stage represents the number of conditions fulfilled in each step. 
\textbf{Engagement Chart.} The engagement chart (Fig.~\ref{fig:line_chart}) shows the efforts students pay on each step when solving the question. It consists of a dual-axis line chart (top and bottom). The upper line chart shows the average time spent in seconds on each step, and the lower line chart shows the trajectory length of the cursor in pixels in each step. \textcolor{black}{The middle blocks between the two line charts indicate the total number of students who tried that particular question and the shaded portions in the $i$-th block represents the number of students who have progressed from $i$-th step to $(i+1)$-th step.} This contextual line chart helps users in understanding how much effort the students make or how much they engage in solving the question based on the time spent and the interactions. It also servers as part of task~\textbf{T4}. When one or more subgroups are selected, a line with a different color is added to show the average time and interaction in each step of that particular group. In Fig.~\ref{fig:line_chart}, the light blue line represents students fulfilling zero conditions and the orange line represents students fulfilling all the six conditions. We can see that students got zero score actually exert more efforts for each step. 

\textbf{Control Panel.} The control panel, as shown in (Fig.~\ref{fig:teaser}(b1)), provides filter options to facilitate exploration. The leftmost filter enables question designers to select a single student and analyze his/her problem-solving path. The middle filter filters the transitions less than a certain number and makes the common path appealing. It also incorporates a navigation bar to provide question designer an overall impression of the length of total steps students take by highlighting the current page using a red rectangle.

\subsubsection{Comparison View}
The Comparison View, as shown in Fig.~\ref{fig:teaser}(c), aims to provide a summarized representation and comparison on how different groups of students' approach a question presented to them. For example, differences in the order of fulfilling the conditions (problem-solving logic), efforts made (engagement) and difficulties encountered by different groups of students could be learned from this view. 

Each group is represented as a column in the Comparison View, such as (Fig.~\ref{fig:teaser}(c1), (c2), and (c3)). Without any selection/filtering from the user, the Comparison View shows only the information of the whole group of students who tried the current question (Fig.~\ref{fig:teaser}(c1)). For each column, the bottom part has a bar chart representing the total number of students who attended the problem along with a box plot that indicates the distribution of the time spent on the problem. Above this part is a series of glyphs stacked vertically up from a lower stage to a higher stage, representing which conditions are satisfied by how many times on each stage.
The unit of the condition glyph in Comparison View is ``time(s)'', which is different from ``student(s)'' in the Transition View.
\textcolor{black}{Given that a student may hit a stage many times at different steps (as shown in the highlighted path in Fig.~\ref{fig:teaser}(b)), and if we only show the number of students who finally hit a stage, the procedure information is lost. Therefore, we summarized the conditions satisfied whenever the group of students hit a certain stage and a darker color represents more times students satisfy the condition, which is different from the transition view. The width of the glyph represents the number of times students reach a particular stage.}
% A darker color represents more times students satisfy the condition, and a wider glyph represents more times students hit that stage.
In addition, box plots are presented between each stage and downward arrow on the right side of each glyph. The box plot represents the average transition time for that particular stage. The size of arrow mark on the side indicates the number of drops/stops on that stage (i.e. the times students dropped to a lower stage or stayed on that particular stage). For example, the larger the size of the arrow, the larger the number of drops/stops in that stage. This visual cue gives users an insight about which stage is most difficult for the students. 

When the user makes a selection/filtering to compare two or more groups, the Comparison View shows cases with similar visualization portraying the information according to the selected group. \textcolor{black}{Fig.~\ref{fig:teaser}(c2) and (c3) show the summarized information of students with zero mark (i.e students fulfilling zero condition, S0) and students with full mark (i.e students fulfilling all six conditions, S6)}. This view also incorporates a legend on the topmost part which aids the users to get a sense of the visual encoding used in the system.

%After the user selecting a group according to the score or grade in the Overview, it adds one column to display the information of the corresponding group. For example, Fig.~\ref{fig:teaser}-c2 represents students who got full mark (fulfilling all six conditions, denoted as 'S6') and Fig.~\ref{fig:teaser}-c3 represents students who got zero mark (fulfilling zero condition, denoted as 'S0')

%% file: sections/06-evaluation.tex
\section{Evaluation}
We evaluate the usefulness and usability of our system through three case studies on real-word datasets (four questions) and interviews with three new domain experts. \textcolor{black}{Their exploration processes generally follow a macro-meso-micro level of analysis while different users may have various tasks and focus, and may not necessarily use all the three analysis levels for one task.}

\subsection{Case Studies}
We report three case studies that were observed by the two question designers (E3 and E6) during their exploration of our system.
The background of the two question designers are introduced in Section~\ref{sec-system-overview}. 
%  \xm{In general, when analyzing a particular group of students, the experts started from the transition view and focused on meso-level and micro-level analysis; when comparing students from different groups, experts started from the overview to the transition view or comparison view and focused on macro-level and meso-level analysis.}
% \yong{The newly-added sentences are not convincing, as no solid arguments are provided.}

\textbf{Check the Gap between Design Intention and Behavior.}
In this case, we describe how our system can help question designers inspect students' problem-solving behaviors and check whether students' problem-solving logic matches the question designer's design intention.
The design intention referred here is reflected in the step-wise solution provided by the question designer.
% We show two examples: one with students whose behavior is similar to the anticipation of question designer, and the other  with students whose behavior deviates from the design intention. 
We will show two question examples found by the domain experts, where the student behaviors either match or deviate from the question designers' design intention.
\begin{figure}[t!]
  \centering 
  \includegraphics[width=0.9\linewidth]{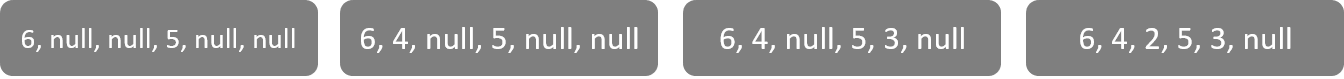}
  \vspace{-0.4cm}
  \caption{A sequence of popular answers.}
   \label{fig:sequence}
\end{figure}

\textbf{\textit{Example 1:}} The question in Fig.~\ref{fig:teaser}(a1) asks students to position the six digits in a way such that the result of its product is as large as possible. E3 thought students may first figure out putting the largest two numbers in the hundred digits and estimated that students may encounter difficulties when dealing with the tenth digits, which was inferred from the post-question solution as shown in Fig.~\ref{fig:example}(b).
% . Thus, E3 provided the post-question solution as shown in Fig.~\ref{fig:example}(b). 
Then, E3 checked the Transition View (Fig.~\ref{fig:teaser}(b2)) to see whether students' practical behaviors met their design intention. He noticed a thick line in the middle of the Sankey diagram as indicated by the orange arrow. This thick line shows that most students got stuck at Stage two and cannot move further to higher stages (i.e., fulfilling more conditions) even after making several steps. Further, by checking the condition glyphs on Stage two, E3 saw that the first and fourth rectangles have darker shade than other rectangles, which represents that these two conditions are fulfilled by more students than other conditions. Referring to the condition description in the top left corner of Fig.~\ref{fig:teaser}(b2), E3 confirmed that the students were clear that the first step is to position the largest numbers (5 and 6) to the hundredths place. He further checked the detailed information shown in the tooltip (Fig.~\ref{fig:sequence}) when hovering the cursor on each step of Stage two. E3 found that the most popular answers given by the students are as follows: \textcolor{black}{$(6, null, null, 5, null, null)$, $(6, 4, null, 5, null, null)$, $(6, 4, null, 5, 3, null)$, $(6, 4, 2, 5, 3, null)$}. It verifies that students were indeed confused in deciding the number for tenths place. In this case, we can safely conclude that the design intention is consistent with students' problem-solving behaviours.

\begin{figure}[t!]
  \centering 
  \includegraphics[width=0.9\linewidth]{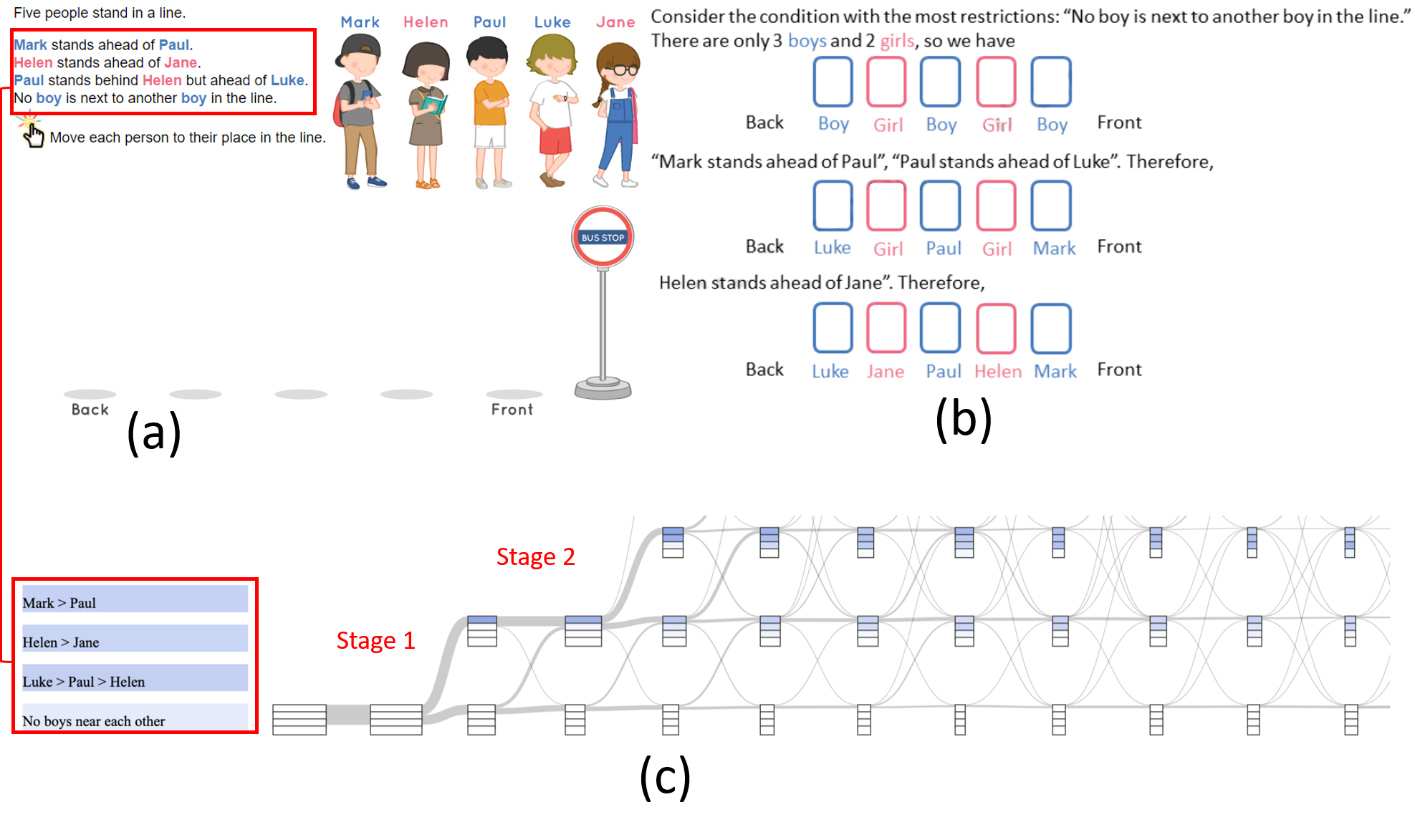}
  \vspace{-0.4cm}
  \caption{Another example of the interactive question (a), its solution (b), and the corresponding transition view (c).}
  \label{fig:example2}
\end{figure}

% \begin{figure}[t!]
%   \centering 
%   \includegraphics[width=1.0\linewidth]{vgtc_journal_latex/pictures/example2_transition.png}
%   \vspace{-0.4cm}
%   \caption{The transition graph of question in Fig.~\ref{fig:example2}.}
%   \label{fig:example2_transition}
% \end{figure}

\textbf{\textit{Example 2:}} The question in Fig.~\ref{fig:example2}(a) requires students to position five different characters in a way such that it satisfies the four conditions described in the question:(1) Mark stands ahead of Paul; (2) Helen stands ahead of Jane; (3) Paul stands behind Helen but ahead of Luke; (4) No boy is next to another boy in the line. This question intends to cultivate students' reverse thinking ability, which means that the students should start solving this problem by fulfilling the last condition first. This thinking logic can achieve a shortest path to solve the problem. Based on this, the provided solution is shown in Fig.~\ref{fig:example2}(b). Then E3 inspected students' problem-solving processes in the Transition View, as shown in Fig.~\ref{fig:example2}(c) and checked whether students have this thinking style. The rectangles in the condition glyphs have the same order with conditions given in the question description as highlighted by the red rectangle. E3 found that all the condition glyphs on Stage one have a darker shade in the first rectangle. In addition, on Stage two, most glyphs have darker shades on the first three rectangles and white in the fourth rectangle. This indicates that all the students started solving the problem by fulfilling the first condition and none started from the last condition. This pattern implies that students may have difficulty coming up with the idea to start from the last condition and also find hard to follow the solution offered. Therefore, this example shows that the design intention, i.e., cultivating the reverse thinking ability, is not matched with students' problem-solving behaviours.

\begin{figure}[t!]
  \centering 
  \includegraphics[width=0.95\linewidth]{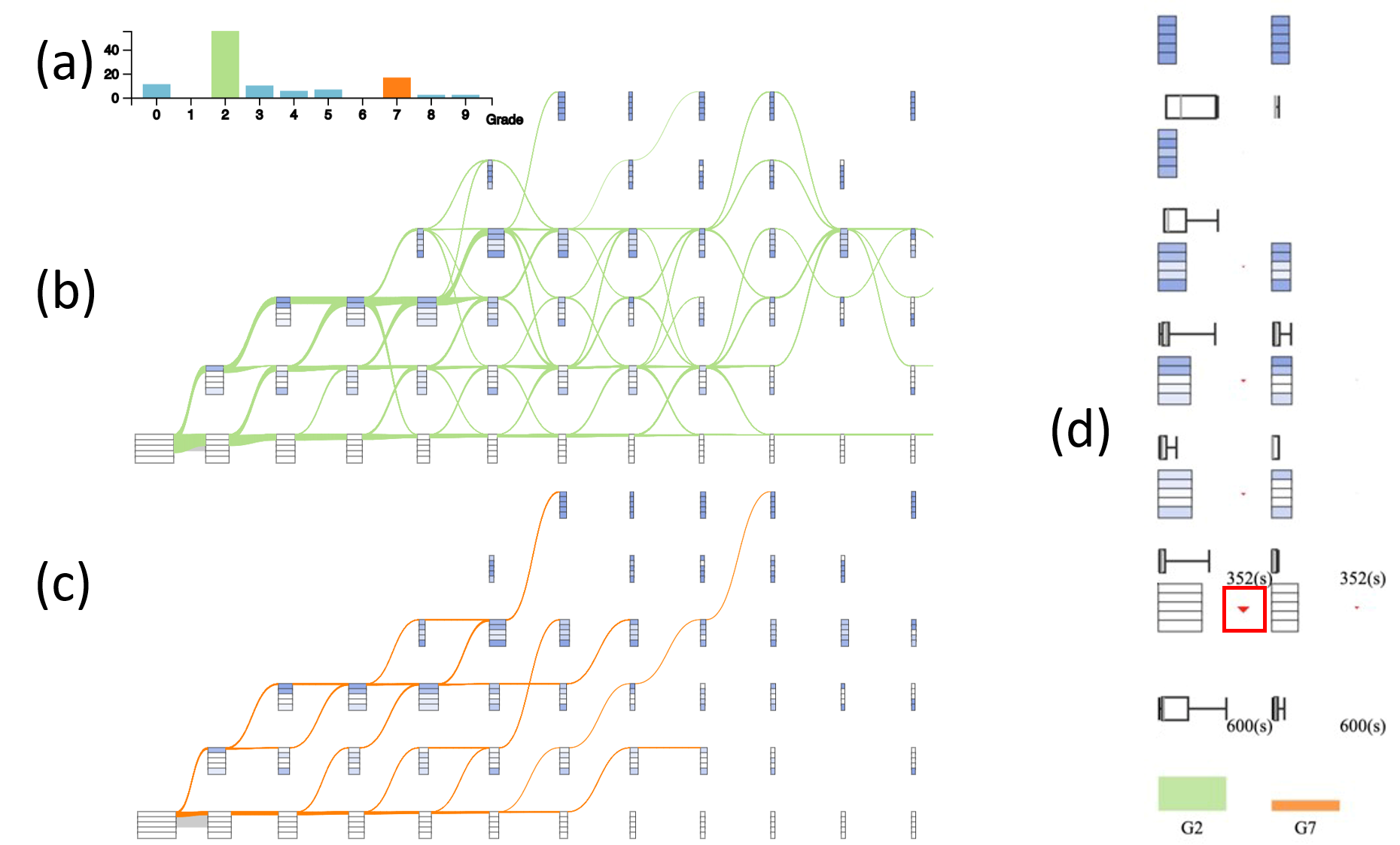}
  \vspace{-0.2cm}
  \caption{The grade distribution of the selected question (a), the transition graphs of students from Grade two (b) and Grade seven (c), and the Comparison View of students from Grade two and Grade seven (d).}
   \label{fig:example3}
\end{figure}

\begin{figure}[t!]
  \centering 
  \includegraphics[width=0.95\linewidth]{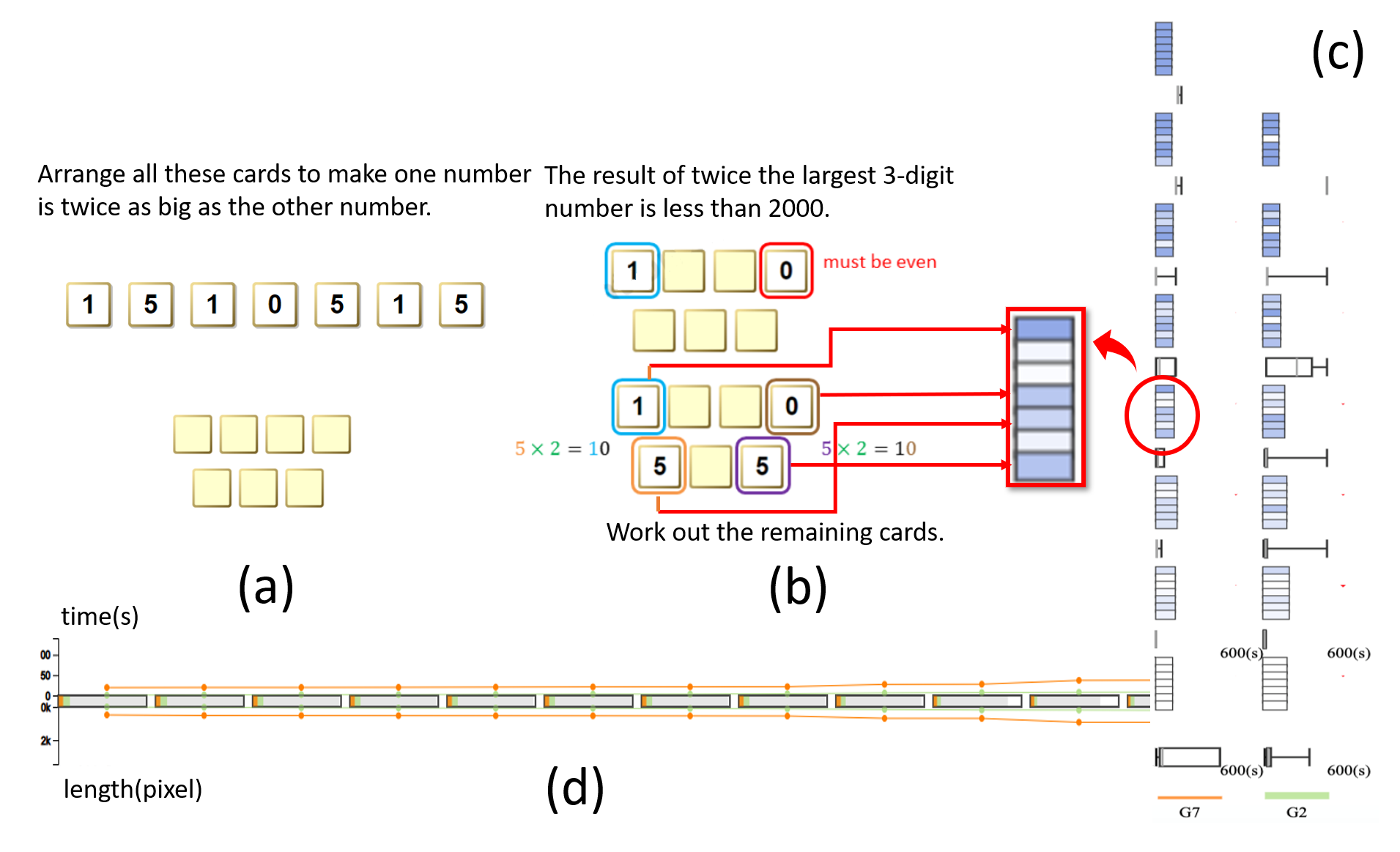}
  \vspace{-0.5cm}
  \caption{Another interactive question (a) and its solution (b), the contextual line chart of students from Grade seven and Grade two (c), and the Comparison View of students from Grade seven and Grade two (d).}
   \label{fig:example4}
\end{figure}

\textbf{Determine the Target Group.}
This case describes that question designers need to compare the problem-solving behaviors of different groups to determine the target group. According to the zone of proximal development (ZPD) principle in education ~\cite{veresov2004zone}, the question should be assigned to a group of students who can interact more with it and do not have major difficulties in solving it.

In this case, the other question designer, E4, analyzed a question that asks students to sort five numbers: $1km, 2500m, 5000mm, 350m, 3km, 500m$
% from the smallest to the largest.
in an ascending order.
From the grade distribution chart (Fig.~\ref{fig:example3}(a)), she found that students from Grade two and Grade seven have tried this problem more than students from other grades. Then, E4 compared the behavior of students from the two grades. In the Transition View of Grade two (Fig.~\ref{fig:example3}(b)), we can see the green lines have lots of ups-and-downs, while the Transition View of students from Grade seven (Fig.~\ref{fig:example3}(b)) have fewer steps and very few transition lines that go down (Fig.~\ref{fig:example3}(c)). Further, E4 compared their problem-solving logic and engagement in detail in the Comparison View (Fig.~\ref{fig:example3}(d)). She quickly noticed that there is an explicit red triangle beside the first stage of Grade two (highlighted with the red rectangle in Fig.~\ref{fig:example3}(d)), which gives her a clue that it is not easy to progress for students from Grade two at the beginning. They check the color distributions of condition glpyhs for these two groups on each stage from the bottom to the top and find that Grade seven has no glyphs on Stage five.
\textcolor{black}{The reason is that some students in Grade seven made a step that fulfilled two of the several conditions required for solving the question. Thus, the Grade seven students directly jumped from Stage four to Stage six.}
% The reason is that students in Grade seven exchanged two elements in their answers on Stage four and the number of conditions satisfied jumped from Stage four to Stage six. 
Based on these observations, E4 believed that the question is difficult for students from Grade two than Grade Seven. However, she found the color distribution of students from Grade two and seven are generally similar, which means they fulfill conditions in a similar order.
%In general, these two groups have a similar color distribution, which means that they fulfilled the conditions in similar order and have similar problem-solving logic.
Based on above explorations and observations, E4 thought that this question can be assigned to a lower grade, such as Grade two, since they can follow the logic and are also engaged in the problem-solving process, while it would be too easy for students from Grade seven.

The question in Fig.~\ref{fig:example4}(a) asks students to place the seven digits such that the upper number is twice as the bottom number. By referring to the solution in Fig.~\ref{fig:example4}(b), the intention of this question is to fix the highest and lowest digits first and then fill the remaining positions. E4 first checked whether students from different grades can catch up with this thinking logic and how much effort they spent on each step. By inspecting the condition glyphs of Grade seven in Comparison View (Fig.~\ref{fig:example4}(c)), E4 found that the first, fourth, fifth, and seventh rectangles become darker from the bottom to the top gradually. Particularly at Stage four, as highlighted by the red circle (Fig.~\ref{fig:example4}(c)), students got these four conditions correct. This indicates that the students in Grade seven followed the design intention to fix the highest and lowest digits at first. However, no such pattern can be found in the condition glyphs of students from Grade two. By comparing the effort students paid on each step in Fig.~\ref{fig:example4}(d), E4 noticed that the orange line (Grade seven) always has a larger value on each step than that of the green line (Grade two) and the green line approaches the x-axis with nearly zero values. This means that students from Grade seven took more time to think for each step while students from Grade two tried the question randomly with less thinking time. Based on these findings, E4 believed that this question may be too difficult for students of a lower grade (e.g., Grade two) since they could not catch up on the design intention.
%idea1 similar score
%idea2 thumb nail picture of transition view can be added to the comparison view

% \begin{figure}[t!]
%   \centering 
%   \includegraphics[width=1.0\linewidth]{vgtc_journal_latex/pictures/example5.png}
%   \vspace{-0.4cm}
%   \caption{(a) Common error panel for a particular question. (b) The typical incorrect path and the data-driven recommended path for the first common error.}
%   \label{fig:example5}
% \end{figure}

% \begin{figure}[t!]
%   \centering 
%   \includegraphics[width=1.0\linewidth]{vgtc_journal_latex/pictures/example5_1.png}
%   \vspace{-0.4cm}
%   \caption{(a) Common error panel for a particular question. (b) Part of the data-driven recommended path for the third common error.}
%   \label{fig:example5-1}
% \end{figure}

\begin{figure}[t!]
  \centering 
  \includegraphics[width=1.0\linewidth]{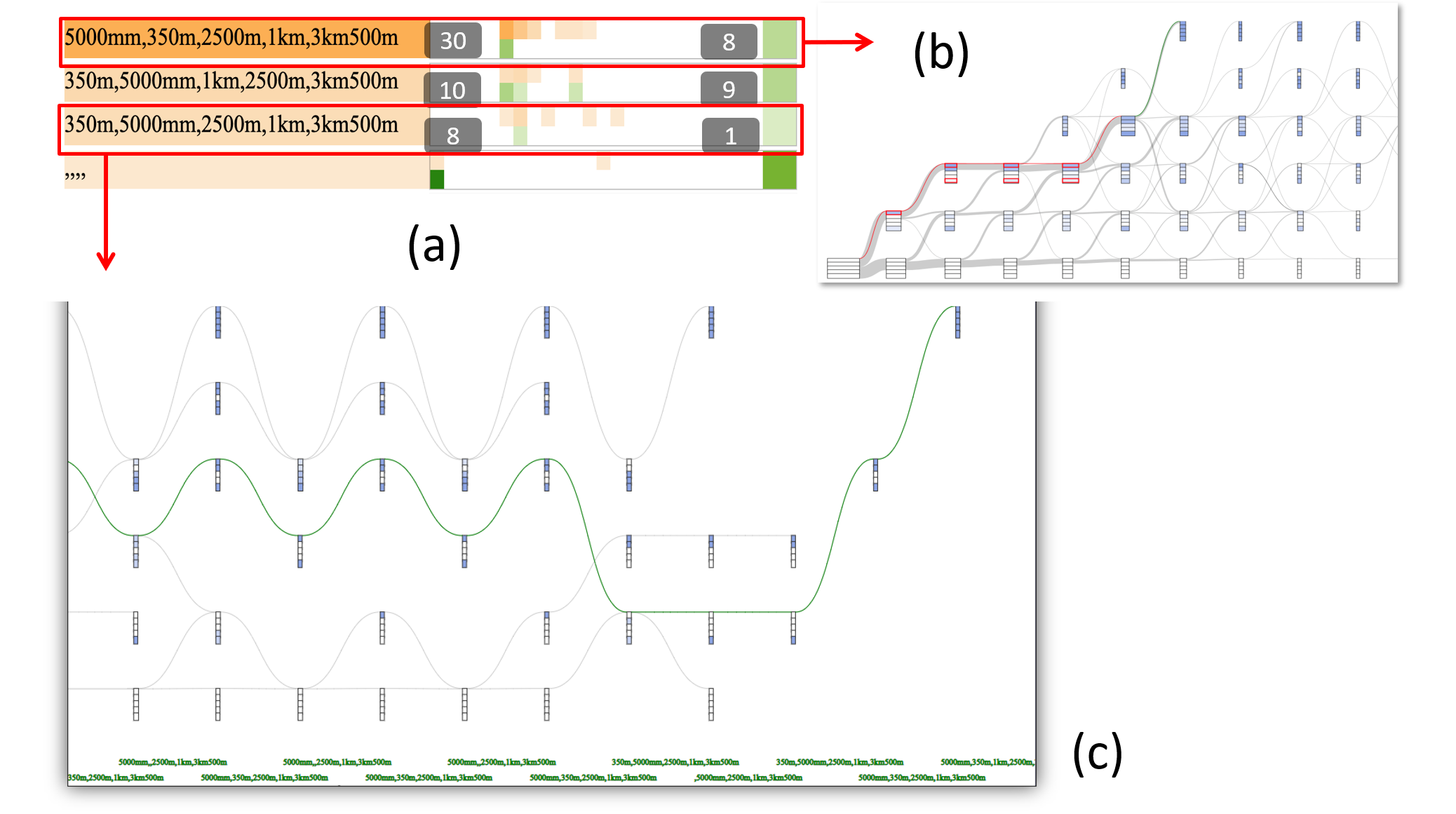}
  \vspace{-0.4cm}
  \caption{Common error panel for a particular question (a), the data-driven recommended path for the first common error (b), one part of the data-driven recommended path for the third common error (c).}
   \label{fig:example5-2}
\end{figure}

\textbf{Evaluate the Feasibility of Data-driven Feedback.}
This case describes how question designers check whether existing data are sufficient to construct and provide data-driven feedback, which is the recommended solution path calculated in Section 4.3. E4 analyzed the same question as in the case ``Determine the Target Group'', which asks students to sort five numbers: $1km, 2500m, 5000mm, 350m, 3km500m$ from the smallest to the largest. E4 first checked the common error panel to see the frequent incorrect answers in Fig.~\ref{fig:example5-2}(a) and found the four most common errors. 
For each of the four errors, E4 checked how many students surpassed the incorrect answers and solved the question in the end. For the first error, 30 students have made mistakes and eight students got a full mark.
% by overcoming the incorrect answer.

Then E4 inspected how this mistake was made and what was the data-driven feedback given by the system. By clicking on the first error, the result is shown in Fig.~\ref{fig:example5-2}(b). The red path in the figure is the incorrect path of that error and the green path is the recommended path based on the existing data. E4 felt that the data-driven feedback for this error is of high quality and reasonable. From the shade of the green squares on the right-hand side, E4 found that not many students have passed the third error, which is indicated by a relatively lighter shade of the square when compared with other squares (Fig.~\ref{fig:example5-2}(a)). 
By clicking on the third error, E4 noticed that the recommended path had a lot of ups and downs as shown in Fig.~\ref{fig:example5-2}(c). The quality of this recommended path was not good and cannot be directly used as feedback for the students. Therefore, by using \textit{QLens}, E4 concluded that current data is insufficient to provide data-driven feedback for all the errors and that some feedback needs to be crafted.

\subsection{Expert Interviews}
Apart from the cases derived together with the four experts during our iterative design process, 
we also interviewed another three domain experts (E5, E6, and E7), who are not the authors of this paper and have not seen \textit{QLens} before the interviews, to further evaluate the effectiveness and usability of \textit{QLens}. 
% our system was introduced to another three domain experts (E5, E6, and E7, who are not the authors of this paper). 
E5 and E6 work in the question design team of an online education company that offers interactive online math questions for primary and secondary school students. E7 works as a senior manager in a popular education company that provides practice questions of all kinds of subjects for K-12 (Grade one to Grade 12) students. E7 is responsible for professional events, such as developing a website for an exam, and also recruitment. The experts are introduced to our system for the first time.. We conducted semi-structured interviews with the experts.
% to evaluate the usefulness and usability (i.e., visual designs and interactions).
% We also collected experts' suggestions for improvements.

\textbf{Procedure}
Each interview had three sections which altogether lasted for 90 minutes. First, we introduced the system and explained various aspects including the purpose of the system, the data we used,
% and how we proceed with the data to show the intermediate steps and stages,
the visualization views available and its functionalities, the visual encoding used in different panels through several examples. 
Then, we show three cases (as introduced in the case studies) to the experts to explain the usability and interactivity of the system and also asked their comments on each case. Third, we invited the experts to spend some time exploring the system and getting themselves familiar with the system. After the exploration part, we ask them several questions regarding visual design, interactive functionality and the overall usability for enhancing the system. We summed up our observations along with the expert’s feedback as follows.

\textbf{System Usefulness}
On a comprehensive note, all the three experts commented that \textit{QLens} is useful.The experts mentioned that the Overview gives them basic knowledge about the question, diversity of the students who attempted it and the difficulty level of the question which can be assessed from the score distribution. This part helped them quickly gain an understanding of the question. All the experts stated that the Transition View is novel and interesting and intuitively conveys the information about the students' problem-solving logic and engagement by representing the trajectory data in various steps and stages. E5 mentioned that the Transition View helps him identify the inconsistency in their design intention and students' current practices and is useful for refining the question to guide students to think in the way question designers intended. In terms of the Comparison View, E5 stated that comparing the problem-solving logic of students from different grades can help the question designers to assign the different questions for students from the appropriate grade. E6 explicitly mentioned that, \textit{``The insights from Transition View will be very useful for the question designer (for example to decide which question is more suitable for which grade students) and the system developer.''} E7 added that~\textit{``As more and more learning activities conducted are online, it was also very useful to compare students from different schools (e.g., international  and local ones) or regions''.} E7 stated that despite the types of questions used in the system are different, our method to summarise the mouse trajectories in steps and stages is very meaningful.

E5 and E6 acknowledged the use of the data-driven method for on-the-fly recommendation guidance for the students. E5 expressed that in the future they would like to incorporate hints to their e-Learning platform in order to help students solve difficult problems. They said the common errors and other summarized related information can help them focus on certain errors. E5 mentioned, \textit{``It would be nice if I can edit the data-driven feedback in the interface when the automatic one is not good.''} E7 said that it would be helpful for him to give guidance to interviewees in the technical interviews.
% \textit{``In our company, we recruit people by assessing their logical thinking skills through algorithmic questions. Since people usually have different solutions, we find it difficult to guide different interviewees and the solution cannot be fixed. I feel that the guidance should be based on each interviewee's problem-solving logic rather than forcing them to understand a fixed answer.''} 
% E3 said that: \textit{``It's difficult to give on-the-fly guidance for different types of questions. I think you are trying to make the machine play the role of a tutor, which is difficult to achieve.''} 
% He suggested that the system can provide ways to compare different data-driven feedback.
He also pointed out a challenge that collecting data might be difficult if the educational platform is accessed using mobile apps. 

\textbf{Visual Designs and Interactions} 
After the experts have spent a considerable amount of time in exploring different functionalities of the system, we collected their feedback on the visual design of the system. Overall, all the experts were satisfied and comfortable in understanding various visual cues and encoding of the system. They also felt that the designs were intuitive. E7 said, \textit{``it is so clear to view the problem-solving process using the visualization like this (Transition View).''}
In terms of the interactions, all the experts appreciated the interactions supported by the system. E5 and E6 praised the smooth interactions to inspect data form different questions, groups, and a particular student. Further, E7 suggested that though the common error panel seems very useful and intuitive, it can be improved in a way that might help a layman understand its functionality (like adding a tooltip). He stated that, \textit{"Visualizations like the common error panel may not be self-explainable, so adding tooltips in the Transition View about its functionality and visual encoding can help users understand its usage."} 

% But E2 expressed that users might need a considerable amount of time to get used to the system as the information delivered by the system can be little overwhelming. 

% He said that: \textit{"The system presented a lot of information and it was a little difficult for him to grasp all those in one go"}. 

% E1 stated that it would be of better use if the output of data-driven recommendation is a hint rather than the solution itself. She said that:\textit{"We are planning to incorporate on-the-fly guidance in the near future to our platform to help students solve difficult problems and the data-driven recommendation in terms of hints would be very useful."} 

%  In addition to this E3 also suggested adding some tooltip to explain the purpose certain icons like filter option, etc. 

% \textbf{Interactions}

%% file: sections/07-discussion.tex
\section{Discussion}
% \yong{we may consider removing the subsubsection title and replace it with some highlighted keywords at the first of each paragraph, which can save space.}

\textbf{Limitation}
Our evaluation demonstrates the effectiveness of ~\textit{QLens}. Nevertheless, there is still space for improvement. (1) Scalability: The number of steps of some problems is larger than 12, which means the whole transition graph cannot be displayed on one page. To mitigate this issue, we add a navigation bar showing the thumbnail image on top of the transition view to achieve focus + context analysis~\cite{bjork2000redefining}. The number of conditions in most questions is less than 10, while if the number of conditions in other application scenarios is more than 15 then the height of one condition rectangle would be small and not easy to observe. (2) Data issue: first, as mobile learning is becoming popular recently, the touch-screen data should also be considered for analysis. We can replace the mouse movement data with the touch-screen data by modifying the trajectory length attribute. Second, the mouse movement data sometimes cannot be converted into the sequence of intermediate answers if the graphical components are overlapped. Additional information is needed for the data conversion, for example, the id of the web component clicked.
\textcolor{black}{
% Third, there might be some 'unexpected behaviors' as outliers that do not reflect students' thinking logic while the large dataset may mitigate this issue.
Third, an underlying assumption of \textit{QLens} is that most students' mouse interactions should reflect their overall thinking logic in their problem-solving process. However, whether all these mouse interactions accurately reflect all the students' thinking logic needs further research, which, however, is beyond the scope of this paper.}
% (3) Authoring tool: the system currently enables question designers to gain insights on how to improve the design while doesn't support them to edit the question directly. It is inconvenient for question designers to switch between the question design interface and the visual analytics system. These factors are not considered currently but are worth being explored in the future.
\textbf{Generalization}
\textcolor{black}{\textit{QLens} is designed for analyzing the problem-solving behaviors of multi-step questions. But its applications are not limited to the online education domain and can also be extended to other application scenarios that involve dense mouse interactions.
For example, when playing video games, different players may have different actions at the same time and game designers need to analyze the common strategies that players will employ or the common difficulties that players are facing in various stages to evaluate the game designs. \textit{QLens} can be easily extended to facilitate an insightful behavior analysis of video game players.
Also, \textit{QLens} can be applied in analyzing the product-browsing behaviors of customers in online shopping platforms (e.g., Amazon, Taobao). By quickly exploring the detailed mouse interactions of customers, data analysts can gain deep insights into customers' online shopping behaviors and design better product recommendation algorithms.}

%% file: sections/08-conclusion.tex
\section{Conclusion and Future Work}
In this paper, we proposed \textit{QLens}, a visual analytics system that assists question designers in exploring students' step-wise problem-solving behaviors in terms of their problem-solving logic, engagement, and also difficulties encountered. The system integrates a novel glyph-embeded Sankey diagram to facilitate the analysis with multiple coordinated views (Overview, Transition View, and Comparison View). The insights learned from the exploring process further guided question designers to improve the design of the question.
% (1) Macro-level: the Overview shows the overall performance achieved by the students;
% (2) Meso-level: the Transition View visualizes the problem-solving processes intuitively to reflect how a group of students proceed along the way and the Comparison View enables question designers to compare the students form different groups;
% (3) Micro-level: incorrect sequences and the corresponding recommended paths are demonstrated for question designers to evaluate the feasibility of the data-driven feedback.
% In addition, \textit{QLens} enables
% rich interactions (e.g., the tooltip and filter) to show detailed information and facilitates in-depth exploration. 
In future work, we plan to extend this system to an authoring tool for question designers to directly edit the question designs and also derive more intuitive visual designs to provide students with on-the-fly guidance.
% incorporate more data-driven feedback \yongnote{algorithms}{} in \textit{QLens}
% % , according to 
% to indicate students' different status and help 
% % design corresponding visualizations for 
% question designers gain a more comprehensive understanding of students.
% % to compare the generated feedback. 
% \yong{I do not quite understand the first future work. Meng, pls further clarify or revise it.}
% Also, it is also interesting to further provide students with on-the-fly guidances by designing intuitive visual designs.